\documentclass[subscriptcorrection,upint,varvw,hyphenate,nolists,nocopyright]{asmejour}

\usepackage{multirow}
\usepackage{threeparttable}
\usepackage[nameinlink]{cleveref}
\usepackage{placeins}
\usepackage{natbib}

\crefname{figure}{Fig.}{Figs.}
\Crefname{figure}{Figure}{Figures}
\crefname{table}{Table}{Tables}
\Crefname{table}{Table}{Tables}
\crefname{equation}{Eq.}{Eqs.}
\Crefname{equation}{Equation}{Equations}
\crefname{section}{Sec.}{Secs.}
\Crefname{section}{Section}{Sections}
\crefname{appendix}{Appendix}{Appendices}
\Crefname{appendix}{Appendix}{Appendices}

\graphicspath{{Figures/}}

\hypersetup{%
  pdfauthor={Yuyou Zhan, Miguel Arana-Catania, Neil Dhir, and Yiguang Li},
  pdftitle={A Forward Model for Route- and Season-Dependent High-Pressure Compressor Efficiency Deterioration in Turbofan Engines},
  pdfkeywords={compressor fouling, particle deposition, forward modelling, efficiency deterioration, performance prediction},
  pdfsubject={Research Paper submitted to the Journal of Turbomachinery}
}

\JourName{Turbomachinery}
\PaperYear{}

\begin{document}

\SetAuthorBlock{Yuyou Zhan\CorrespondingAuthor}{%
Centre for Propulsion and Thermal Power Engineering,\\
Faculty of Engineering and Applied Sciences,\\
Cranfield University,\\
Cranfield, Bedfordshire MK43 0AL, United Kingdom\\
email: yuyou.zhan@cranfield.ac.uk}

\SetAuthorBlock{Miguel Arana-Catania}{%
Digital Scholarship at Oxford,\\
University of Oxford,\\
Oxford, OX1 3BG, United Kingdom\\
email: humd0244@ox.ac.uk}

\SetAuthorBlock{Neil Dhir}{%
Faculty of Engineering and Applied Sciences,\\
Cranfield University,\\
Cranfield, Bedfordshire MK43 0AL, United Kingdom\\
email: neil.dhir@cranfield.ac.uk}

\SetAuthorBlock{Yiguang Li}{%
Fellow ASME\\
Centre for Propulsion and Thermal Power Engineering,\\
Faculty of Engineering and Applied Sciences,\\
Building 52, Cranfield University,\\
Cranfield, Bedfordshire MK43 0AL, United Kingdom\\
email: i.y.li@cranfield.ac.uk}

\title{A Forward Model for Route- and Season-Dependent High-Pressure Compressor Efficiency Deterioration in Turbofan Engines}


\begin{abstract}
Civil turbofan engines lose compressor efficiency every time they fly through particle-laden air. Fouling of the high-pressure compressor (HPC) is the dominant recoverable performance-loss mechanism. Existing studies treat environmental particle exposure, blade-row deposition and stage-performance deterioration as separate problems, which limits understanding of the issue. Stage-stacking approaches require deterioration levels to be prescribed rather than derived from the operating environment, leaving the causal chain from route conditions to engine performance unresolved. This paper presents a forward simulation framework that closes this gap by propagating route-, season-, and altitude-dependent particle exposure through four successive layers: (1) flight-phase-resolved HPC inlet dose calculation; (2) stage-wise deposition across the HPC; (3) deposition-to-deterioration mapping; and (4) stage thermodynamic stacking, yielding overall fouled HPC isentropic efficiency. Demonstrated on a representative civil high-bypass turbofan baseline with characteristics similar to the CFM56-7B, the predicted deterioration milestones are confirmed in magnitude and timescale against publicly available in-service support datasets. The results reproduce front-stage-dominated deposition, rapid early deterioration with asymptotic saturation, and clear route- and season-dependent fouling behaviour. To our knowledge, this is the first model that closes the complete chain from operationally resolved environmental exposure to the overall loss of HPC isentropic efficiency in a single physically traceable forward framework.
\end{abstract}

\date{}
\maketitle

\section{Introduction}\label{sec:intro}

Civil aero-engines continuously ingest airborne particulate matter during normal operation. Ground manoeuvres, take-off, climb-out, descent, and landing expose the core flow path to dust, sand, sea salt, industrial pollutants, and other atmospheric contaminants whose concentration varies with airport location, season, and altitude \citep{Bojdo2020,Ryder2024,Igie2018} (see \cref{Overall_framework}). Repeated exposure causes particles smaller than approximately 10\,$\mu$m to deposit on compressor blades, vanes, and end walls \citep{Kurz2012,Doring2017a}. The resulting increase in surface roughness and passage blockage increases aerodynamic losses and reduces the static-pressure rise across individual blade rows \citep{Bons2010,Gbadebo2004}. For the high-pressure compressor, these effects accumulate in multiple stages, progressively reducing the overall isentropic efficiency \citep{Doring2017b}. The operational consequences include increased fuel burn, reduced exhaust gas temperature margin, shortened on-wing life, and higher maintenance costs \citep{Richardson1979,Sallee1980}. The cold-section refurbishment of the JT9D fleet, for example, recovered approximately 1.3 percentage points of thrust-specific fuel consumption that had been lost due to compressor degradation \citep{Richardson1979}. The ability to predict how different operating environments affect the progression of fouling and loss of HPC efficiency has direct value for engine health management, maintenance planning, and compressor washing decisions \citep{Igie2018,Suman2017review}.

Previous work has developed prediction models for individual aspects of compressor fouling, including deposition experiments, stage-stacking performance calculations, and dust ingestion estimation \citep{Doring2017a,Lakshminarasimha1986,Bojdo2020}. However, none of these approaches provides a single framework that propagates environment-dependent particle exposure through stage-wise deposition to overall HPC efficiency loss. This paper presents a complete forward simulation framework and demonstrates its application across multiple operating environments. The remainder of this section reviews the relevant literature in detail before presenting the specific contributions of this work.

\subsection{Related Work}

Previous studies have investigated compressor fouling through several complementary approaches. In-service engine programmes, particularly the NASA JT9D Jet Engine Diagnostics Programme, established that HPC performance deterioration increases with flight cycles, with the majority of recoverable losses attributable to surface roughness and contour changes in the compressor \citep{Richardson1979,Sallee1980}. At the blade-row level, cascade and rotating-rig experiments have clarified that deposition is an asymptotic process governed by the balance between particle adhesion and removal, with moisture playing a decisive role in sticking probability \citep{Doring2017a,Vulpio2021rotorcraft}. These experiments also quantified the time scales and magnitudes of the increase in total pressure loss and the reduction in static pressure rise as functions of the mass of ingested particles \citep{Doring2017a}. At the multistage level, compressor fouling campaigns on rotating rigs have confirmed that deposition is concentrated in the front stages and that the fouling intensity depends on humidity and exposure time \citep{Vulpio2021rotorcraft,Igie2017}. Surface roughness studies in cascades and single-stage compressors have further shown that loss increases with roughness and Reynolds number, with the suction-side trailing region being most sensitive \citep{Bammert1972,Bons2010,Back2012}. In parallel, stage-stacking techniques have been used to propagate assumed levels of stage deterioration to overall compressor performance. Early approaches by Lakshminarasimha and Saravanamuttoo \citep{Lakshminarasimha1986} and subsequent refinements by Muir et al.~\cite{Muir1989} and Zaita et al.~\cite{Zaita1998} employed generalised stage characteristics with prescribed efficiency and flow reductions. More recently, D\"{o}ring et al.~\cite{Doring2017b} combined experimentally calibrated blade-row deterioration parameters with a modified stage-stacking procedure to compute deteriorated compressor maps. On the environmental side, Bojdo et al.~\cite{Bojdo2020} and Ryder et al.~\cite{Ryder2024} used reanalysis data to estimate altitude- and season-dependent dust doses at various airports, highlighting the strong spatial and temporal variability of particle exposure.

Despite these advances, existing approaches address different parts of the deterioration chain in isolation. In-service data provide fleet-level trends but rarely contain traceable environmental exposure or stage-level deposition states \citep{Richardson1979,Sallee1980}. Blade-row experiments and test-rig campaigns reveal local mechanisms under controlled conditions but do not directly answer how a given HPC degrades over thousands of flight cycles in a specific operating environment \citep{Doring2017a,Vulpio2021rotorcraft}. Dust ingestion models estimate the particle dose entering the engine but stop short of predicting stage-wise deposition and its effect on HPC efficiency \citep{Bojdo2020,Ryder2024}. Stage-stacking models can propagate a given level of stage deterioration to overall compressor performance but require the deterioration to be prescribed in advance, without generating it from environmental exposure \citep{Lakshminarasimha1986,Doring2017b,Yang2014}. Consequently, a gap remains between local fouling physics and operationally resolved deterioration prediction: no single framework currently propagates route-, season-, and altitude-dependent particle exposure through HPC deposition to overall isentropic efficiency loss in a manner that is compact enough for repeated scenario evaluation, while remaining physically interpretable from input environment to output performance.

\begin{figure*}[!htp]
    \centering
    \includegraphics[width=0.95\textwidth]{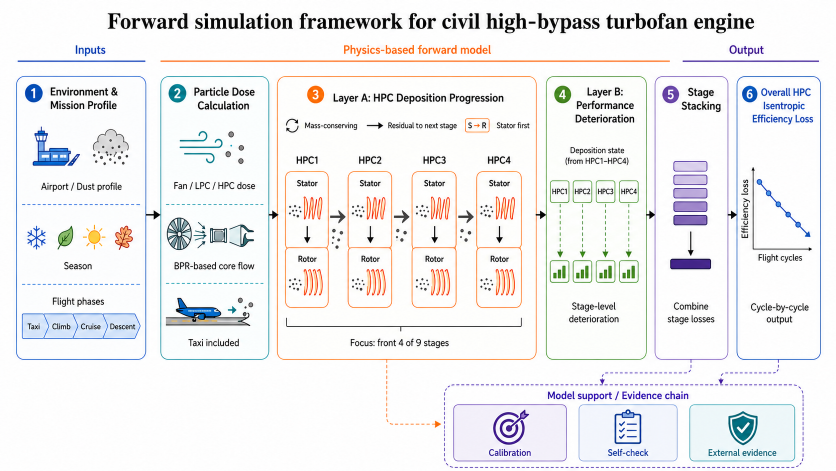}
    \caption{Overall framework of the proposed route- and season-resolved forward simulation chain. The model propagates environmental particle exposure (left) through particle dose calculation, mass-conserving HPC deposition across four front stages, stage-level deterioration mapping, and stage-stacking thermodynamic propagation to predict overall HPC isentropic efficiency loss (right).}
    \label{Overall_framework}
\end{figure*}

To address this need, the present paper proposes a forward simulation framework that links environment-resolved particle exposure, stage-wise HPC deposition, and overall HPC isentropic efficiency deterioration for a representative civil high-bypass turbofan engine. The framework consists of four successive modelling layers:
\begin{enumerate}
\item The first layer integrates airport-, season-, and altitude-dependent particle concentration along the flight profile to obtain the HPC inlet particle dose for each flight leg.
\item The second layer distributes the dose across the front stages, tracking stator and rotor deposition \citep{Doring2017a}.
\item The third layer maps the normalised deposition to stage deterioration correction factors \citep{Doring2017b}.
\item The fourth layer propagates total temperature and pressure through the nine-stage HPC, using the stage coefficients obtained in the previous layer, to obtain the overall fouled isentropic efficiency.
\end{enumerate}
The framework is demonstrated on a representative civil high-bypass turbofan baseline as a case study, with predicted deterioration milestones compared against publicly available in-service support data \citep{Richardson1979,Sallee1980}. This paper makes the following contributions:
\begin{enumerate}
\item We propose a novel forward simulation framework that propagates route-, season-, and altitude-dependent environmental particle exposure through stage-wise HPC deposition to overall isentropic efficiency deterioration, without prescribing any deterioration level in advance.
\item We demonstrate the validity of the model by comparing predicted deterioration milestones against publicly available in-service support data from high-bypass turbofan fleets \citep{Richardson1979,Sallee1980}, showing agreement in both magnitude and timescale.
\item We present that the model produces physically interpretable deterioration trends, including front-stage-dominated deposition and rapid early deterioration followed by asymptotic saturation.
\item We showcase the application of the model on a representative civil high-bypass turbofan baseline across multiple proxy environments spanning clean, moderate, and polluted conditions, confirming that the framework captures route- and season-dependent fouling behaviour.
\end{enumerate}
The overall framework of this paper is shown in \cref{Overall_framework}.

\section{Model Setup, Operating Scenarios and Data Sources}\label{sec:setup}

\subsection{Engine Description}\label{sec:baseline}

\begin{figure}[!htb]
    \centering
    \includegraphics[width=\columnwidth]{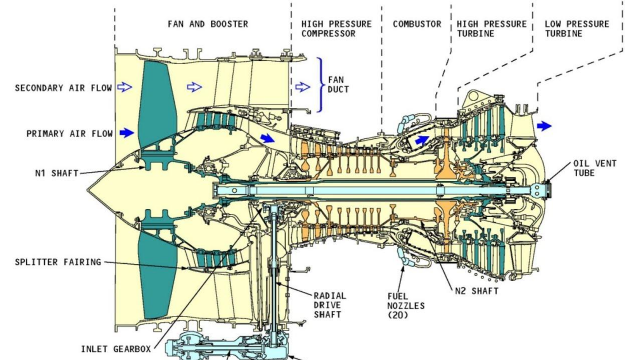}
    \caption{Cross-sectional schematic of a representative two-spool high-bypass turbofan engine, showing the fan, low-pressure compressor, nine-stage high-pressure compressor, and other components. The primary (core) and secondary (bypass) air flow paths are indicated. Adapted from~\cite{ScienceDirectJetEngine}.}
    \label{fig2}
\end{figure}

The framework is demonstrated on a representative civil high-bypass turbofan engine with thermodynamic and geometric characteristics similar to those of the CFM56-7B. The engine configuration comprises a single-stage wide-chord fan, a three-stage low-pressure compressor (LPC), a nine-stage high-pressure compressor (HPC), a double-annular combustion chamber, a single-stage high-pressure turbine, and a four-stage low-pressure turbine \citep{AircraftCommerce2008,DeltaTechOps}. A cross-sectional schematic is shown in \cref{fig2}. The key engine-level parameters used in this study are summarised in \cref{tab:engine_params} \citep{AircraftCommerce2008,Turan2022,Halliwell2022,Dviation2020}.

\begin{table}[!htp]
\caption{Key engine-level parameters of the representative civil high-bypass turbofan baseline.}
\label{tab:engine_params}
\centering
\begin{tabular*}{\columnwidth}{@{\extracolsep\fill}lr@{}}
\toprule
Parameter & Value \\
\midrule
Overall pressure ratio (OPR) & 32.7 \\
Bypass ratio (BPR) & 5.3:1 \\
Fan pressure ratio (FPR) & 1.60:1 \\
LPC pressure ratio & $\approx$2.00 \\
HPC pressure ratio & $\approx$10.5 \\
Total mass flow rate & 341\,kg\,s$^{-1}$ \\
HPC stage count & 9 \\
\bottomrule
\end{tabular*}
\end{table}

The stage-level geometric parameters required by the deposition model, namely blade chord length $c_j$, blade span $h_j$, and aspect ratio $\mathrm{AR}_j = h_j / c_j$, are assembled from publicly available sources for a nine-stage HPC representative of civil high-bypass turbofan practice. The HPC inlet and exit hub and casing radii reported by Halliwell and Watsek~\cite{Halliwell2022} are used as boundary conditions, and the intermediate stage-wise radii are estimated by linear interpolation. The blade span at each stage is then computed from the interpolated hub and casing radii. The chord length is derived from the aspect ratio, which is assumed to decrease linearly from inlet to exit in a manner consistent with published design practice \citep{Grieb2009}. The blade counts for the first four stator and rotor rows are obtained from CFM56-family training and borescope-inspection manuals \citep{Dviation2020,Jiang2004}. The complete stage-wise geometry is listed in \cref{tab:depcap}.

The resulting geometric parameters are cross-validated against several independent sources. The Grieb turbomachinery textbook \citep{Grieb2009} tabulates typical ranges of chord length (30--50\,mm), aspect ratio (1.0--2.0), and hub-to-tip ratio (0.6--0.9) for high-bypass turbofan HPC stages; all values used in this study fall within these ranges. D\"{o}ring et al.~\cite{Doring2017a} report chord length, aspect ratio, solidity, hub-to-tip ratio, and stage pressure ratio for a representative HPC cascade, and the present values are consistent with that dataset. The aspect ratio and hub-to-tip ratio are further confirmed against the ranges given by Farokhi~\cite{Farokhi2014}, Boyce~\cite{Boyce2011}, Wu et al.~\cite{Wu1950}, and \"{U}st\"{u}nda\u{g} et al.~\cite{Ustundag2018}.

Each HPC stage is characterised by three normalised performance coefficients at the clean (unfouled) condition: the temperature coefficient $\zeta_{0,j}$, the isentropic pressure-rise coefficient $\psi_{0,j}$, and the isentropic efficiency $\eta_{0,j}$, defined as
\begin{align}
\zeta_{0,j} &= \frac{\Delta h_{t,j}}{U_j^2/2} \label{eq:clean_zeta} \\
\psi_{0,j} &= \frac{\Delta h_{t,s,j}}{U_j^2/2} \label{eq:clean_psi} \\
\eta_{0,j} &= \frac{\psi_{0,j}}{\zeta_{0,j}} \label{eq:clean_eta}
\end{align}
where $U_j$ is the blade speed at mean radius, and $\Delta h_{t,j}$ and $\Delta h_{t,s,j}$ are the actual and isentropic specific enthalpy rises across the stage, respectively \citep{Doring2017b}. These coefficients describe the baseline aerodynamic performance before any particle deposition occurs; their fouled counterparts are derived in \cref{sec:stacking}. The construction procedure, which applies dual global scaling factors to shape priors resampled from the NASA Energy Efficient Engine dataset, is documented in Appendix~\ref{app:stageparams}. The resulting nine-stage clean baseline values are listed in \cref{tab:cleanbaseline}. Asymptotic deterioration correction factors $\chi_{\eta,\infty,j}$ and $\chi_{\zeta,\infty,j}$ are initialised from experimentally determined cascade data \citep{Doring2017b} and distributed across the affected stages following a linear spatial model, and the values used are presented in \cref{tab:chi}.

\subsection{Deposition Capacity, Fouling-Affected Domain, and Rotor-to-Stator Ratio}\label{sec:depcap}

\begin{figure*}[!htp]
    \centering
    \includegraphics[width=0.95\textwidth]{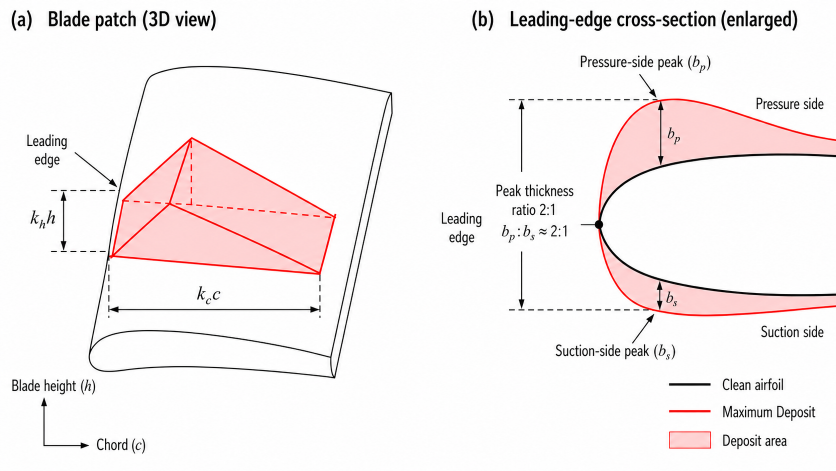}
    \caption{Geometric assumptions for estimating the single-blade maximum deposition capacity. (a) Three-dimensional view of the blade patch, showing the effective chord $k_c c$ and effective span $k_h h$ that define the deposition area. (b) Enlarged leading-edge cross-section, showing the pressure-side peak deposit thickness $b_p = 450\,\mu$m and suction-side peak thickness $b_s = 225\,\mu$m, with a pressure-to-suction thickness ratio of approximately 2:1.}
    \label{fig:geometric_assumption}
\end{figure*}

The maximum deposition mass that a single blade can accumulate is estimated from a simplified deposition volume. The geometric assumptions used in this estimation are presented in \cref{fig:geometric_assumption}. For a blade type $b$ (stator or rotor) at stage $j$, we estimate the saturation mass from the simplified deposited volume as:
\begin{align}
\label{eq:Minf}
M_{\infty,b,j} &= \frac{1}{2}\rho_p \, V_{\mathrm{dep},j} \nonumber \\
&= \frac{1}{2} \rho_p \; k_c \, c_j \; k_h \, h_j \; \delta_{\mathrm{layer}} 
\end{align}
where $\rho_p = 2700$\,kg\,m$^{-3}$ is the deposit layer density corresponding to typical aluminosilicate mineral particles \citep{Vulpio2022, ISO12103_1_2024}, $c_j$ (m) is the geometric chord length, $h_j$ (m) is the blade span, $\delta_{\mathrm{layer}}$ (m) is the deposit layer thickness, and $k_c \in (0,1]$ and $k_h \in (0,1]$ are chordwise and spanwise coverage factors that account for the non-uniform spatial distribution of deposition on blade surfaces \citep{Mullaney2025,Kurz2012}. The effective geometric parameters $k_c c_j$ and $k_h h_j$ are determined through calibration, as described in \cref{sec:calibration_procedure}, and their numerical values are listed in \cref{tab:depcap}. The deposit thickness $\delta_{\mathrm{layer}}$ is set by combining pressure-side and suction-side representative thicknesses. The pressure-side deposit thickness is set to $b_p = 450\,\mu$m based on experimental measurements reported by Mullaney et al.~\cite{Mullaney2025} and Syverud et al.~\cite{Syverud2007}. Experimental studies consistently indicate that deposit thickness on the pressure surface is approximately twice that on the suction surface \citep{Tang2020,Casari2020,Vulpio2021rotorcraft}. Accordingly, the suction-side thickness is set to $b_s = b_p / 2 = 225\,\mu$m. In this study, therefore, the deposit thickness $\delta_{\mathrm{layer}} = (b_p + b_s)$.

The deposition-active domain is restricted to the first four HPC stages. Multiple experimental studies confirm that compressor fouling is concentrated in the front stages, where inflow particle concentration is highest and upstream blade rows progressively deplete the available cumulative dose \citep{Doring2017b,Suman2017review,Zhao2025}. The choice of four stages represents a reasonable first-order assumption for the present HPC configuration, and the remaining five stages participate in the thermodynamic stacking calculation but are treated as clean.

Since the asymptotic deposition kernel used in this framework is derived from stator cascade experiments \citep{Doring2017a}, it cannot be applied directly to rotor blades for which equivalent experimental time-scale data are not available. To handle rotor deposition without introducing an uncalibrated dynamic model, the framework constrains rotor mass through a rotor-to-stator deposition ratio $\gamma_{rs}$. Multistage fouling experiments show that stator blades accumulate roughly 2 times more deposit mass than rotor blades at the same stage, owing to differences in centrifugal loading, relative velocity, and wall shear stress \citep{Zhao2025,Vulpio2021rotorcraft,Tarabrin1998}. The ratio $\gamma_{rs}$ is calibrated within this experimentally supported range, and its optimal value and the calibration procedure are documented in \cref{tab:posterior} and \cref{sec:calibration_procedure}.

\subsection{Operating Scenarios, Flight Profiles, and Environmental Data}\label{sec:scenarios}

The particle concentration entering the model, $C_p$, is determined by airport $a$, season $s$, and altitude $z$, so that $C_p = C_p(a,s,z)$. Altitude- and season-dependent dust concentration profiles are derived from the Copernicus Atmosphere Monitoring Service (CAMS) reanalysis dataset collected and processed by Ryder et al.~\cite{Ryder2024} for global airports. Each flight leg is discretised into sequential phases with altitude, Mach number, and low-pressure spool speed ($N_1$) specified for each segment. The take-off profile comprises five phases: taxi-out at ground level, take-off thrust preset, take-off roll with initial rotation, initial climb under take-off/go-around thrust to approximately 1,600\,m, and continued climb to 6,000\,m. The descent profile comprises four phases: high-altitude descent from 6,000\,m to 1,000\,m, approach with deceleration, final descent to touchdown, and taxi-in. The altitude, time, Mach number, and $N_1$ ranges for each phase are summarised in \cref{tab:flight_profile}, based on published landing and take-off (LTO) cycle standards \citep{ICAO2010}, flight-test and emission data \citep{He2024,Graver2009,Prakash2016}, and the altitude profile convention adopted by Ryder et al.~\cite{Ryder2024}. Taxi durations at each airport are taken from the Eurocontrol taxi-time statistical reports \citep{Eurocontrol2024}. Engine mass flow rates across flight phases are simulated using the Pythia gas-turbine performance simulation tool \citep{Li2014Pythia}. No additional engine parameters beyond those presented in this paper are required to reproduce these results.

\begin{table*}[!t]
\caption{Flight-profile phases used for mass-flow simulation. Taxi durations are airport-specific and obtained from Eurocontrol~\cite{Eurocontrol2024}.}\label{tab:flight_profile}
\centering
\begin{tabular*}{\textwidth}{@{\extracolsep\fill}llllll@{}}
\toprule
Segment & Phase & Time (s) & Altitude (m) & Mach & Low-pressure Spool Speed $N_1$ (\%) \\
\midrule
\multirow{5}{*}{Take-off}
& Taxi-out & Airport-specific & 0 & 0 & 30 \\
& Initial climb & 30--150 & 20--1,600 & 0.22--0.44 & 80--92 \\
& Climb to 6,000\,m & 150--600 & 1,600--6,000 & 0.44--0.62 & 92--95 \\
\midrule
\multirow{4}{*}{Landing}
& High-alt. descent & 0--540 & 6,000--1,000 & 0.65--0.24 & 65--50 \\
& Approach + final & 540--780 & 1,000--0 & 0.24--0.22 & 50--25 \\
& Taxi-in & Airport-specific & 0 & 0 & 30 \\
\bottomrule
\end{tabular*}
\end{table*}

\begin{figure}[!htb]
    \centering
    \includegraphics[width=\columnwidth]{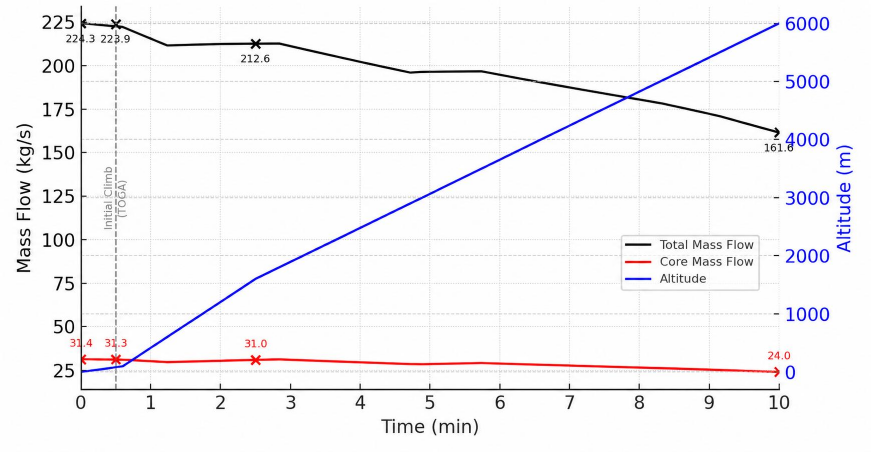}
    \caption{Simulated total and core mass flow rates during the take-off and climb phase (0--6,000\,m), obtained from the Pythia gas-turbine performance tool \citep{Li2014Pythia}. The altitude profile is shown on the right axis (blue). Mass flow decreases with altitude as air density drops, while the core-to-total flow split is governed by the bypass ratio. These profiles serve as input to the per-flight HPC inlet dose calculation (see \cref{eq:dose}).}
    \label{fig:Takeoff}
\end{figure}

\begin{figure}[!htb]
    \centering
    \includegraphics[width=\columnwidth]{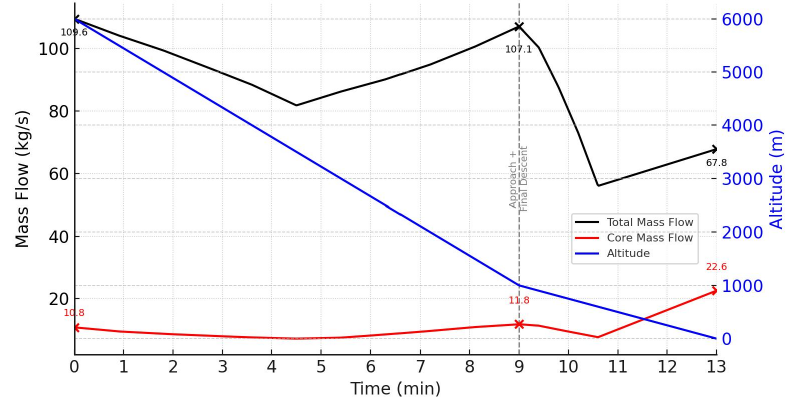}
    \caption{Simulated total and core mass flow rates during the descent phase (6\,000--0\,m). The altitude profile is shown on the right axis. The descent mass flow is lower than the climb phase due to reduced thrust settings. Together with \cref{fig:Takeoff}, these profiles define the complete flight-leg mass flow input to the dose model.}
    \label{fig:Descent}
\end{figure}

The simulation results are shown in \cref{fig:Takeoff,fig:Descent}. Different airports and seasons produce different per-flight HPC inlet doses, enabling the model to respond to route- and season-dependent exposure rather than assuming a fixed contaminant concentration. The operating scenarios considered in this study include a calibration anchor route and several ordinary in-service proxy routes spanning a range of environmental severity. The details are provided in \cref{sec:results}.

\subsection{Engine Flow Path and Particle Transmission}\label{sec:transmission}

\begin{figure*}[!t]
    \centering
    \includegraphics[width=0.95\textwidth]{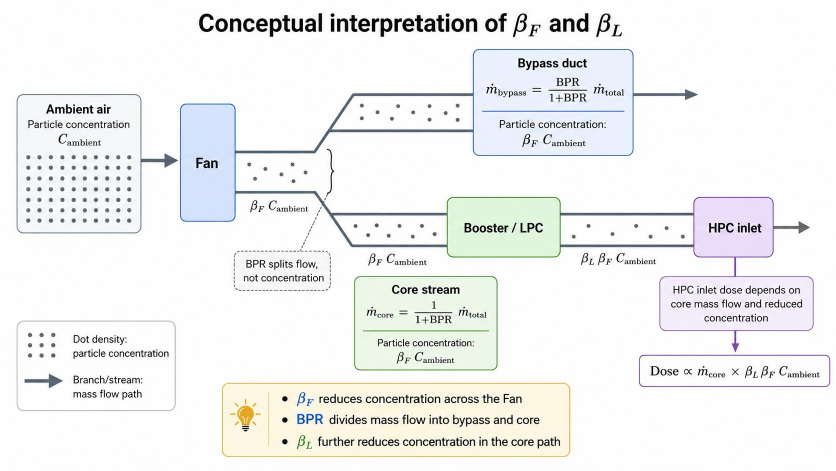}
    \caption{Conceptual interpretation of the particle transmission factors $\beta_F$ and $\beta_L$. Ambient air with particle concentration $C_{\mathrm{ambient}}$ enters the fan (left). The fan reduces the concentration to $\beta_F \, C_{\mathrm{ambient}}$ through particle-blade interactions; this reduced concentration applies to both the bypass and core streams. The bypass ratio splits the total mass flow between the bypass duct and the core stream but does not alter the particle concentration. The booster/LPC further reduces the concentration in the core path to $\beta_L \, \beta_F \, C_{\mathrm{ambient}}$ at the HPC inlet (right). Dot density represents particle concentration; arrow represents mass flow rate.}
    \label{fig:beta_schematic}
\end{figure*}

Particles ingested at the fan inlet undergo two successive concentration reductions before reaching the HPC, as illustrated in \cref{fig:beta_schematic}. It is important to distinguish these concentration reductions from the mass flow split caused by the bypass ratio, because the two mechanisms play fundamentally different roles.

The fan-to-core transmission factor $\beta_F$ represents a reduction in particle mass concentration across the fan disk. As ambient air passes through the fan, particle-fan interactions such as inertial impaction and centrifugal separation cause a fraction of particles to deposit on fan blade surfaces or to be ejected from the main flow. The particle concentration downstream of the fan is therefore reduced to $\beta_F \, C_{\mathrm{ambient}}$, where $C_{\mathrm{ambient}}$ is the ambient particle mass concentration. This reduced concentration applies uniformly to the entire post-fan air stream. Particle-fan interaction simulations indicate that $\beta_F \in [0.78,\,0.86]$ \citep{Vogel2019}.

The bypass ratio then divides the total mass flow into two streams: the bypass duct carries $\mathrm{BPR}/(1+\mathrm{BPR})$ of the total mass flow, while the core stream carries $1/(1+\mathrm{BPR})$. This split affects only the mass flow rate, not the particle concentration. Both the bypass and core streams carry the same reduced concentration $\beta_F \, C_{\mathrm{ambient}}$.

The LPC-to-HPC transmission factor $\beta_L$ represents a second concentration reduction as the core-stream particles pass through the booster stages. Particle deposition within the LPC further depletes the particle loading, so that the concentration at the HPC inlet becomes $\beta_L \, \beta_F \, C_{\mathrm{ambient}}$. Multistage fouling experiments suggest that $\beta_L$ depends on relative humidity (RH), with $\beta_L \in [0.77,\,0.984]$ at 50\,\%\,RH and $\beta_L \in [0.72,\,0.984]$ at 80\,\%\,RH \citep{Vulpio2021rotorcraft,Vulpio2021timewise}.

The per-flight HPC inlet dose (\cref{eq:dose}) therefore integrates the product of core volumetric flow rate $\dot{m}_{\mathrm{core}}/\rho_{\mathrm{air}}$ and the doubly reduced concentration $\beta_L \, \beta_F \, C_{\mathrm{ambient}}$ over all flight-profile segments. Both $\beta_F$ and $\beta_L$ are calibrated within the experimentally supported prior ranges listed above, according to the procedure of \cref{sec:calibration_procedure}; their optimal values are reported in \cref{tab:params} and the calibration search ranges are listed in \cref{tab:posterior}.

\subsection{Scope and Assumptions}\label{sec:scope}

The following assumptions define the scope and boundaries of the present model. They are stated explicitly so that the reader can assess the applicability of the results and identify directions for future refinement.

\begin{enumerate}
\item The model predicts overall HPC isentropic efficiency deterioration only. It does not compute engine-level thrust, specific fuel consumption, exhaust gas temperature, or full compressor maps.
\item Deposition dynamics are modelled explicitly for the first four HPC stages. The remaining five stages participate in the stage-stacking thermodynamic calculation but are not subject to deposition-induced deterioration.
\item Fan and LPC deposition are not modelled as independent dynamic processes. Their net particle retention effect is represented by two lumped concentration weakening factors, $\beta_F$ and $\beta_L$ (see \cref{sec:transmission}).
\item The HPC operating point, defined by the corrected mass flow and corrected speed at the compressor inlet, is not re-solved as fouling progresses. The stage-stacking calculation uses the design-point flow coefficient $\phi^{*} = 1$ throughout.
\item The core mass flow profile is not dynamically updated in response to fouling. It remains fixed at the clean-engine values obtained from the Pythia simulation (\cref{sec:scenarios}).
\item The overall HPC pressure ratio is not treated as an independent deterioration variable. Changes in pressure ratio arise solely as natural outputs of the stage-stacking propagation with fouled stage coefficients.
\item In the proxy-environment simulations, the aircraft operates return flights within a single airport environment to isolate the effect of environmental severity on fouling rate. This does not imply that real airline operations are restricted to a single environment.
\end{enumerate}

\section{Methodology}\label{sec:method}

The proposed framework (see \cref{Overall_framework}) propagates environment-resolved particle exposure through four successive modelling layers: HPC inlet dose calculation, stage-wise deposition, deposition-to-deterioration mapping, and stage-stacking thermodynamic propagation. Each layer takes the output of the preceding one as its sole input, forming a unidirectional forward chain that terminates at the overall HPC isentropic efficiency. The following subsections describe each layer in turn.

\subsection{HPC Inlet Particle Dose}\label{sec:dose}

For the $k$-th flight leg, the particle mass entering the HPC inlet is computed by integrating the product of core volumetric flow rate and local particle concentration over all flight-profile segments. We define the per-flight HPC inlet cumulative dose as:

\begin{equation}\label{eq:dose}
C_{\mathrm{HPC,in}}^{(k)}
= \beta_F \, \beta_L \sum_{i} \frac{\dot{m}_{\mathrm{core},i}}{\rho_{\mathrm{air},i}} \, C_{p}(a,s,z_i) \, \Delta t_i 
\end{equation}

where $C_{p}(a,s,z_i)$ (kg\,m$^{-3}$) is the particulate mass concentration determined by airport $a$, season $s$, and altitude $z_i$; $\dot{m}_{\mathrm{core},i} = \dot{m}_{\mathrm{total},i}/(1+\mathrm{BPR})$ is the core mass flow rate (kg\,s$^{-1}$); $\rho_{\mathrm{air},i}$ (kg\,m$^{-3}$) is the International Standard Atmosphere (ISA) air density at $z_i$; and $\Delta t_i$ (s) is the segment duration. The coefficients $\beta_F \in [0, 1]$ and $\beta_L \in [0, 1]$ represent the fan-to-core and LPC-to-HPC particle transmission factors, respectively, introduced in \cref{sec:transmission}.

\subsection{Stage-Wise Deposition}\label{sec:deposition}

Particles entering the HPC are distributed across the first four stages. Our model calculates the cumulative dose in each stage sequentially, considering the mass conservation between stages. Let $C_j^{(k)}$ denote the cumulative dose available to stage $j$ during flight $k$, with $C_1^{(k)} = C_{\mathrm{HPC,in}}^{(k)}$. Within each stage, the stator row deposition is calculated first, using the result to calculate the rotor deposition. In the following equations, $m_{s,j}^{(k)}$ denotes the accepted per-blade stator deposit mass at stage $j$ after flight $k$, which satisfies both the deposition model and the mass-conservation constraint. The superscript \textit{raw} indicates the unconstrained output of the exponential deposition kernel before the mass-conservation clip is applied, and the superscript $\textit{capped}$ indicates the geometrically capped rotor mass before the residual-dose clip is applied. Following the asymptotic Kern--Seaton deposition kernel \citep{Doring2017a}, the mass deposit increase over time considers inhibitive mechanisms (e.g., autoretardation, particle re-entrainment) proportional to the mass already deposited, and thus behaves according to the following equation:

\begin{equation}
\begin{aligned}
\frac{dm}{dt} &= R_0 - R_1 m \\
&\Rightarrow m(t) = \frac{R_0}{R_1} - \left(\frac{R_0}{R_1} - m(t=0)\right)\exp(-R_1 t)
\end{aligned}
\end{equation}

where $R_0$ is the deposition rate at the beginning of the deposition process, and $R_1$ represents the inhibitive mechanisms. 

Transforming the equation from the time domain to the mass domain according to \citep{Doring2017a}, and substituting for our scenario, the single-blade stator mass after flight $k$ is
\begin{equation}\label{eq:stator}
m_{s,j}^{\mathrm{raw},(k)}
= M_{\infty,s,j}
- \bigl(M_{\infty,s,j} - m_{s,j}^{(k-1)}\bigr)
\exp\!\Bigl(-\frac{C_j^{(k)}}{n_{s,j}\,\tau}\Bigr) 
\end{equation}
where $M_{\infty,s,j}$ (g\,blade$^{-1}$) is the stator saturation mass estimated from a simplified blade-surface deposition volume, $n_{s,j}$ is the stator blade count, and $\tau$ (g) is the mass-domain deposition time constant calibrated from cascade experiments \citep{Doring2017a}. The total stator deposition increment is then clipped against the available dose to enforce mass conservation: 
\begin{equation}\label{eq:clip}
\Delta M_{s,j}^{(k)} = \min\!\left(n_{s,j}(m_{s,j}^{\mathrm{raw},(k)} - m_{s,j}^{(k-1)}),\; C_j^{(k)}\right)
\end{equation}
Rotor deposition is not driven by the exponential kernel, since $\tau$ originates from stator cascade evidence. Instead, rotor mass is constrained by a rotor-to-stator deposition ratio $\gamma_{rs}$, whose value has been introduced in \cref{sec:depcap}, and a geometric cap:
\begin{equation}\label{eq:rotor}
m_{r,j}^{\mathrm{capped},(k)}
= \min\!\left(\gamma_{rs}\, m_{s,j}^{(k)},\; M_{\infty,r,j}\right)
\end{equation}
where $M_{\infty,r,j}$ is the rotor saturation mass. The rotor increment is likewise clipped against the residual dose after stator deposition. The cumulative dose passed to the next stage satisfies
\begin{equation}\label{eq:throughdose}
C_{j+1}^{(k)} = C_j^{(k)} - \Delta M_{s,j}^{(k)} - \Delta M_{r,j}^{(k)}
\end{equation}
which ensures that particles captured upstream are never double-counted downstream. We define the stage normalised deposition $f_j^{(k)}$, which represents the ratio of the current total deposit mass at stage $j$ to the model-consistent saturation mass and serves as the input to the next modelling layer:
\begin{equation}\label{eq:fprogress}
f_j^{(k)} = \frac{n_{s,j}\,m_{s,j}^{(k)} + n_{r,j}\,m_{r,j}^{(k)}}{M_{\infty,j}^{\mathrm{model}}}
\end{equation}
where $M_{\infty,j}^{\mathrm{model}} = n_{s,j}\,M_{\infty,s,j} + n_{r,j}\min(\gamma_{rs}\,M_{\infty,s,j},\,M_{\infty,r,j})$ is the model-consistent saturation mass and $f_j^{(k)} \in [0,1]$.

\subsection{Deterioration Mapping}\label{sec:bridge}

In the stage-stacking framework of D\"{o}ring et al.~\cite{Doring2017b}, the deterioration correction factors $\chi_{\eta}$ and $\chi_{\zeta}$ are applied directly at their asymptotic values corresponding to full saturation (Eqs.~13 and~15 in that reference). To model the intermediate states between clean and fully fouled conditions, we introduce a normalised deterioration driver $d_j^{(k)}$ that scales linearly between zero (clean) and one (fully saturated). We propose to connect the previously calculated normalised deposition $f_j$ to $d_j$ through a power-law function:
\begin{equation}\label{eq:bridge}
d_j^{(k)} = \bigl(f_j^{(k)}\bigr)^{\beta^{*}}
\end{equation}
where $\beta^{*}$ is a shape exponent whose optimal value is obtained through the calibration procedure described in \cref{sec:calibration_procedure}. 

Two stage-level correction factors, the efficiency modifier $\chi_{\eta,j}$ and the loading modifier $\chi_{\zeta,j}$, are then obtained as
\begin{align}
\chi_{\eta,j}^{(k)} &= 1 + (\chi_{\eta,\infty,j} - 1)\,d_j^{(k)} \label{eq:chi_eta} \\
\chi_{\zeta,j}^{(k)} &= 1 + (\chi_{\zeta,\infty,j} - 1)\,d_j^{(k)} \label{eq:chi_zeta}
\end{align}
where the asymptotic values $\chi_{\eta,\infty,j}$ and $\chi_{\zeta,\infty,j}$ represent the correction factors at full saturation. These expressions reduce to the clean condition ($\chi = 1$) when $d_j = 0$ and recover the full asymptotic values of D\"{o}ring et al.~\cite{Doring2017b} when $d_j = 1$. The first-stage values are taken from experimentally determined cascade data \citep{Doring2017b}. The asymptotic correction factors are distributed across the $K$ affected stages using a linear spatial model. For stages $j = 1, \ldots, K$, the first-stage values $\chi_{\eta,\infty,1}$ and $\chi_{\zeta,\infty,1}$ are taken from the experimental cascade data of D\"{o}ring et al.~\cite{Doring2017b}, and the values decrease linearly to unity at stage $K+1$.
\begin{align}
\chi_{\eta,\infty,j} &= 1 + (\chi_{\eta,\infty,1} - 1)\,\frac{K + 1 - j}{K} \label{eq:chi_spatial_eta} \\
\chi_{\zeta,\infty,j} &= 1 + (\chi_{\zeta,\infty,1} - 1)\,\frac{K + 1 - j}{K} \label{eq:chi_spatial_zeta}
\end{align}
where $j = 1, \ldots, K$.
Stages $j > K$ are unaffected and retain $\chi_{\eta,\infty,j} = \chi_{\zeta,\infty,j} = 1$. In this study, $K = 4$ and $\chi_{\eta,\infty,1} = 1.093$, $\chi_{\zeta,\infty,1} = 1.023$ \citep{Doring2017b}. The resulting stage-wise values are listed in \cref{tab:chi}.

\subsection{Stage Stacking and Overall HPC Isentropic Efficiency}\label{sec:stacking}

Following D\"{o}ring et al.~\cite{Doring2017b}, each compressor stage is characterised by three normalised coefficients: temperature coefficient (\cref{eq:clean_zeta}), isentropic pressure-rise coefficient (\cref{eq:clean_psi}), and isentropic efficiency (\cref{eq:clean_eta}). The clean baseline values $\zeta_{0,j}$, $\psi_{0,j}$, and $\eta_{0,j}$ are obtained by scaling a set of shape priors $\tilde{\zeta}_j$ and $\tilde{\eta}_j$, resampled from the ten-stage NASA Energy Efficient Engine dataset \citep{Doring2017b}, using two global factors $a$ and $b$. The factor $a$ scales the efficiency priors so that $\eta_{0,j} = a\,\tilde{\eta}_j$, while the factor $b$ scales the temperature coefficient priors so that $\zeta_{0,j} = b\,\tilde{\zeta}_j$. The pressure-rise coefficient follows as $\psi_{0,j} = \zeta_{0,j}\,\eta_{0,j}$. The values of $a$ and $b$ are determined simultaneously by requiring that the resulting stage-stacked HPC satisfies two closure conditions: (i) the total stagnation temperature rise matches the target overall HPC temperature rise, and (ii) the overall pressure ratio $\prod_{j=1}^{9} \pi_{0,j}$ matches the target HPC pressure ratio. This procedure yields $a = 0.9055$ and $b = 0.8856$, and the complete derivation is provided in Appendix~\ref{app:stageparams}. Fouled normalised coefficients are computed from the correction factors using the formulation of D\"{o}ring et al.~\cite{Doring2017b}:

The fouled normalised efficiency is
\begin{equation}\label{eq:fouled_eta}
{\eta_j^{\prime}}^{*} = C_{\eta,j} - (C_{\eta,j} - 1)\,\chi_{\eta,j},
\end{equation}
where $C_{\eta,j} = 1/\eta_{0,j}$. The fouled normalised temperature coefficient is
\begin{equation}\label{eq:fouled_zeta}
{\zeta_j^{\prime}}^{*} = C_{\zeta,j} - (C_{\zeta,j} - 1)\,\chi_{\zeta,j},
\end{equation}
where $C_{\zeta,j} = 2/\zeta_{0,j}$. All deterioration corrections are evaluated at the fixed clean design operating point. Therefore, the normalised flow coefficient is fixed at
\begin{equation}\label{eq:flow_coefficient}
\phi_j^*=\frac{\phi_j}{\phi_{0,j}}=1.
\end{equation}
The relation $\eta_j^*=1$ applies only to the clean reference state. Under fouled conditions, the normalised coefficients $\eta_j'{}^*$ and $\zeta_j'{}^*$ are calculated from \cref{eq:fouled_eta} and \cref{eq:fouled_zeta}, respectively. The present model does not
re-solve the shift in flow coefficient caused by fouling. The fouled normalised pressure-rise coefficient follows as
\begin{equation}\label{eq:fouled_psi}
{\psi_j^{\prime}}^{*} = {\zeta_j^{\prime}}^{*}\,{\eta_j^{\prime}}^{*}.
\end{equation}

We further assume that the isentropic efficiency and flow coefficients are the design values $\eta=\eta_0$ and $\phi=\phi_0$. Dimensional values are recovered by $\zeta_j^{\prime} = \zeta_{0,j}\,{\zeta_j^{\prime}}^{*}$, $\eta_j^{\prime} = \eta_{0,j}\,{\eta_j^{\prime}}^{*}$, and $\psi_j^{\prime} = \psi_{0,j}\,{\psi_j^{\prime}}^{*}$ as listed in \cref{tab:cleanbaseline}. 

The stage thermodynamic propagation converts the fouled normalised coefficients back to dimensional total temperature and pressure changes.

An ideal gas following an isentropic (adiabatic reversible) process follows:
\begin{equation}
    \begin{split}
        PV^\gamma&=\mathrm{constant}\\ 
        P^{1-\gamma}T^\gamma&=\mathrm{constant}
    \end{split}
\end{equation}
which in our case leads to the following relation between the stagnation pressure ratio across a stage and the stagnation temperature ratio:
\begin{equation}\label{eq:isentropic_relation}
\frac{p_{t,j+1}}{p_{t,j}} = \left(\frac{T_{t,j+1}}{T_{t,j}}\right)^{\!\gamma_a/(\gamma_a - 1)}
\end{equation}

The specific enthalpy of an ideal gas follows the relations:
\begin{equation}
    \begin{split}
    dh &= c_p dT\\
    \gamma&=\frac{c_p}{c_v}
    \end{split}
\end{equation}
where $c_p$ and $c_v$ are the specific heats at constant pressure and constant volume, respectively, and which using \cref{eq:clean_zeta} leads to:
\begin{equation}\label{eq:Tout}
T_{t,j+1}^{\prime} = T_{t,j}^{\prime} + \frac{\zeta_j^{\prime}\,U_j^2/2}{2c_p}
\end{equation}
where $T^{\prime}$ refers to a temperature value in the fouled scenario.

Defining the stage pressure ratio as $\pi_j^{\prime}=p_{\mathrm{out},j}^{\prime}/p_{\mathrm{in},j}^{\prime}$ and using \cref{eq:isentropic_relation} and \cref{eq:Tout} leads to the following equation for the stage pressure ratio in terms of the fouled pressure-rise coefficient:
\begin{equation}
\pi_j^{\prime} = \Bigl(1 + \frac{\psi_j^{\prime}\,U_j^2/2}{c_p\,T_{t,j}^{\prime}}\Bigr)^{\frac{\gamma_a}{\gamma_a - 1}}
\end{equation} 

After propagating through all nine stages, the overall fouled HPC isentropic efficiency is
\begin{equation}\label{eq:etaHPC}
\eta_{\mathrm{HPC,is}}^{\prime(k)}
= \frac{c_p\,T_{t,25}\Bigl[\bigl(p_{t,3}^{\prime(k)}/p_{t,25}\bigr)^{(\gamma_a-1)/\gamma_a} - 1\Bigr]}
{\displaystyle\sum_{j=1}^{9} \Delta h_{t,j}^{\prime(k)}}
\end{equation}
where $T_{t,25}$ and $p_{t,25}$ are the HPC inlet total temperature and pressure, and $p_{t,3}^{\prime(k)} = p_{t,25}\prod_{j=1}^{9}\pi_j^{\prime(k)}$ is the fouled HPC exit total pressure, with subscripts 25 and 3 denoting the HPC inlet and exit stations following standard engine station numbering convention. The deterioration after $k$ flights is quantified as $\Delta\eta_{\mathrm{HPC,is}}^{(k)} = \eta_{\mathrm{HPC,is}}^{\prime(k)} - \eta_{\mathrm{HPC,is},0}$.

\section{Results and Discussion}\label{sec:results}

\subsection{Calibrated Fixed-Parameter Setup}\label{sec:paramsetup}

Our model contains several physically interpretable but not directly measurable parameters, including the particle transmission factors $\beta_F$ and $\beta_L$, the characteristic mass scale of the deposition kernel $\tau$, the chordwise and spanwise coverage factors $k_c$ and $k_h$, the rotor-to-stator deposition ratio $\gamma_{rs}$, and the bridge exponent $\beta^{*}$ governing the nonlinear mapping from deposition progress to deterioration. The deposition layer parameters ($\tau$, $k_c$, $k_h$, $\beta_F$, $\beta_L$, $\gamma_{rs}$) are calibrated against the cascade deposition experiments of D\"{o}ring et al.~\cite{Doring2017a} under a fixed ground-level particle concentration of $48\,\mu\mathrm{g\,m}^{-3}$, matching the Barcelona reference condition. The calibration procedure is described in \cref{sec:calibration_procedure}. The bridge exponent $\beta^{*}$ is calibrated separately at the performance layer against the stage-stacking deterioration results of D\"{o}ring et al.~\cite{Doring2017b}, using the same ground-level concentration with an exponentially decaying altitude profile and no seasonal variation.

Once optimised, the same parameter set is used for all route- and season-resolved simulations reported below. No parameter needs to be adjusted for individual airports, seasons, or output curves. Differences among the reported scenarios therefore arise entirely from the operating environment, the mission profile, and the forward propagation of the model, rather than from case-specific tuning. The optimal values used for all subsequent simulations are summarised in \cref{tab:params}. Once determined, these values are held fixed for all subsequent simulations. The fact that no parameter is adjusted between scenarios means that all differences in the reported results arise solely from the environmental input and mission profile.

\begin{table*}[!htp]
\caption{Fixed and calibrated model parameters used for all simulations reported in this study.}\label{tab:params}
\centering
\begin{threeparttable}
\setlength{\tabcolsep}{8pt}
\begin{tabular*}{0.90\textwidth}{@{\extracolsep\fill}llll@{}}
\toprule
Symbol & Description & Optimal value & Unit \\
\midrule
$\tau$ & Mass-domain deposition constant & 3.7 & g \\
$k_c$ & Chordwise coverage factor & 0.12 & -- \\
$k_h$ & Spanwise coverage factor & 0.30 & -- \\
$\beta_F$ & Fan-to-core transmission factor & 0.78 & -- \\
$\beta_L$ & LPC-to-HPC transmission factor & 0.85 & -- \\
$\gamma_{rs}$ & Rotor-to-stator deposition ratio & 0.48 & -- \\
$\rho_p$ & Deposit layer density & 2,700 & kg\,m$^{-3}$ \\
$\beta^{*}$ & Power-law bridge exponent & 0.94 & -- \\
\bottomrule
\end{tabular*}
\begin{tablenotes}[flushleft]
\footnotesize
\item Stage-wise geometry, clean-stage coefficient distributions, closure factors, deterioration modifier arrays, and calibration search ranges are provided in Appendices~\ref{app:depcap}--\ref{app:posterior}.
\end{tablenotes}
\end{threeparttable}
\end{table*}

\subsection{Parameter Calibration Procedure}\label{sec:calibration_procedure}

The model parameters listed in \cref{tab:params} are obtained through a two-layer calibration procedure. In the first layer, the deposition model parameters are calibrated against the cascade deposition experiments of D\"{o}ring et al.~\cite{Doring2017a}. The calibration uses a fixed ground-level particle concentration of $48\,\mu\mathrm{g\,m}^{-3}$, matching the Barcelona reference condition reported in that study. The calibration proceeds in three sequential stages. In Stage~A, the chordwise and spanwise coverage factors $k_c$ and $k_h$ are jointly searched over a grid within the ranges listed in \cref{tab:posterior}, with the objective of matching the predicted $N_{50}$, $N_{75}$, and $N_{90}$ milestone (number of annualised flight cycles at which the normalised deterioration $d$ reaches a certain percentage of its asymptotic value) to the experimentally reported value of approximately 247 taxi/idle-equivalent cycles \citep{Doring2017a}. Once $k_c$ and $k_h$ are fixed, Stage~B searches the fan-to-core and LPC-to-HPC transmission factors $\beta_F$ and $\beta_L$ over their respective prior ranges (\cref{tab:posterior}), using the same $N_{xx}$ target. In Stage~C, the rotor-to-stator deposition ratio $\gamma_{rs}$ is searched within the experimentally supported range, again targeting $N_{xx}$ consistency. At each stage, all previously determined parameters are held fixed, ensuring that the calibration is sequential and non-circular. The search is conducted by exhaustive grid evaluation; no gradient-based or stochastic optimiser is used.

In the second layer, the bridge exponent $\beta^{*}$ is calibrated separately against the stage-stacking deterioration results of D\"{o}ring et al.~\cite{Doring2017b}. The particle concentration profile uses the same ground-level value of $48\,\mu\mathrm{g\,m}^{-3}$ with an exponentially decaying altitude dependence and no seasonal variation. The objective function is a weighted logarithmic error between the predicted and reference deterioration trajectories across the full cycle range. The search is conducted over the range $\beta^{*} \in [0.50,\;1.00]$ at a resolution of 0.01. The optimal value $\beta^{*} = 0.94$ minimises the weighted log-error. The complete set of search ranges and optimal values is summarised in \cref{tab:posterior}.

Once all parameters are determined, the complete parameter set is fixed, and no further adjustment is made for any subsequent simulation. The distinction between the two calibration layers reflects the modular structure of the framework: the deposition layer parameters are calibrated against deposition data alone, while the bridge exponent is calibrated against performance deterioration data, ensuring that each parameter is informed by the most relevant experimental evidence.

\subsection{Model Magnitude Validation}\label{sec:credibility}

Before interpreting route- and season-dependent behaviour, it is necessary to validate the predicted deterioration magnitude. Using the optimal parameters in \cref{tab:params} with the default scenario (low-moisture deposition, $K = 4$ affected stages, full-stage deterioration modifiers from \cref{tab:chi}), and computing the stage-stacked overall efficiency from \cref{eq:etaHPC} with no deposition ($f_j = 0$ for all $j$) and at full saturation ($f_j = 1$ for all $j$), the clean overall HPC isentropic efficiency is 0.901, consistent with the design-point value specified in the baseline dataset \citep{Halliwell2022}. The asymptotic reduction in overall HPC isentropic efficiency is 0.00325, corresponding to 0.325 percentage points or approximately 0.36\% relative to the clean value. The surface-roughness contribution inferred from the in-service data of Richardson et al.~\cite{Richardson1979} is approximately 0.23--0.24 percentage points. The present prediction is therefore moderately higher, but remains of the same order of magnitude. Because the engine type, deterioration definition, operating history, and environmental exposure differ, this comparison is interpreted as an external magnitude check rather than a strict upper bound or engine-specific validation.

\begin{figure*}[!t]
    \centering
    \includegraphics[width=\linewidth]{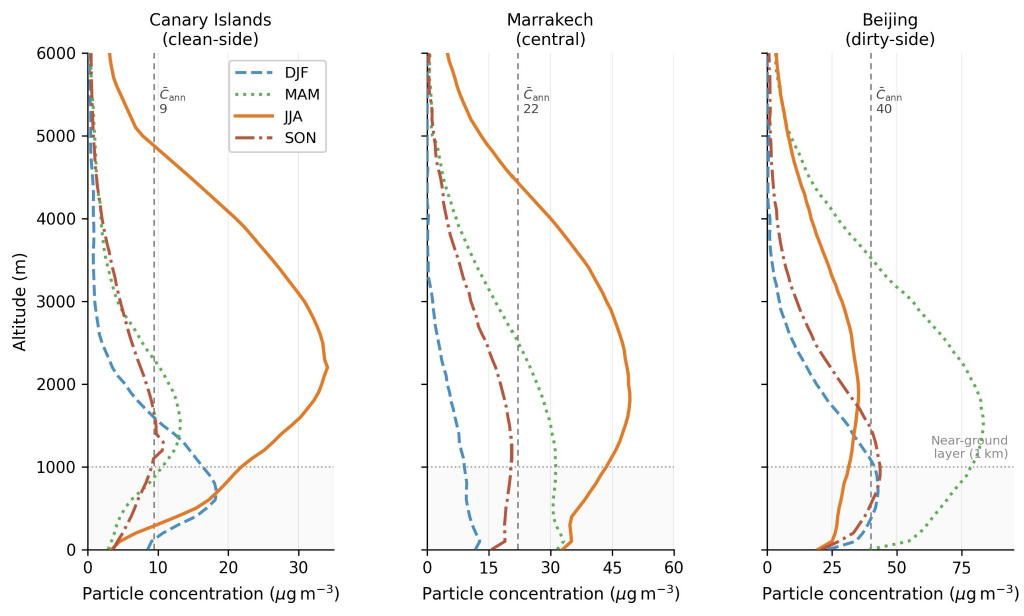}
    \caption{Altitude- and season-resolved particle concentration profiles for the three proxy sites derived from the CAMS reanalysis dataset of Ryder et al.~\cite{Ryder2024}. Each panel shows the four seasonal profiles (DJF, MAM, JJA, SON) for one site. The horizontal dotted line marks the 1,000\,m near-ground layer boundary used to compute the annual geometric-mean concentrations reported in the text. The vertical dashed line in each panel indicates the corresponding annual geometric mean $\bar{C}_{\mathrm{ann}}$ over the 0--1,000\,m layer.}
    \label{fig:profiles}
\end{figure*}

To assess the deterioration timescale, the model is evaluated under three proxy environments that span the range of particle exposure typically encountered in ordinary civil aviation operations, using the altitude-dependent concentration profiles described in \cref{sec:scenarios}. The Canary Islands represent a clean-side proxy, with an annual 
geometric-mean particle concentration of approximately 
$9\,\mu\mathrm{g\,m}^{-3}$ averaged over the 0--1\,000\,m near-ground layer. Marrakech represents a central proxy at approximately $22\,\mu\mathrm{g\,m}^{-3}$, reflecting moderate 
North African dust activity. Beijing represents a dirty-side proxy at approximately $40\,\mu\mathrm{g\,m}^{-3}$, roughly four times the Canary Islands level. These concentrations are derived from the Copernicus Atmosphere Monitoring Service (CAMS) reanalysis profiles of Ryder et al.~\cite{Ryder2024} by computing the 
altitude-averaged geometric mean across four seasons (December--February (DJF), March--May (MAM), June--August (JJA), September--November (SON)) over the 0--1\,000\,m layer, which encompasses the taxi, initial climb, and approach phases where particle ingestion is highest. \cref{fig:profiles} presents the full altitude- and season-resolved concentration profiles for the three proxy sites. The vertical structure differs markedly between locations: the Canary Islands display a near-surface peak that is most pronounced in DJF, consistent with long-range Saharan dust transport arriving at low altitudes during winter; Marrakech shows the highest concentrations in JJA and MAM, with a profile that intensifies above 500\,m in summer, reflecting elevated dust layers driven by convective uplift; Beijing reaches its seasonal maximum in MAM at altitudes of 1,000--2,000\,m, driven by long-range transport from the Gobi and Taklamakan deserts under strong spring cyclone winds \citep{Ryder2024}. These site-specific vertical and seasonal structures are the direct physical input that differentiates the per-flight HPC inlet dose across routes and seasons.

\begin{figure}[!htb]
    \centering
    \includegraphics[width=\columnwidth]{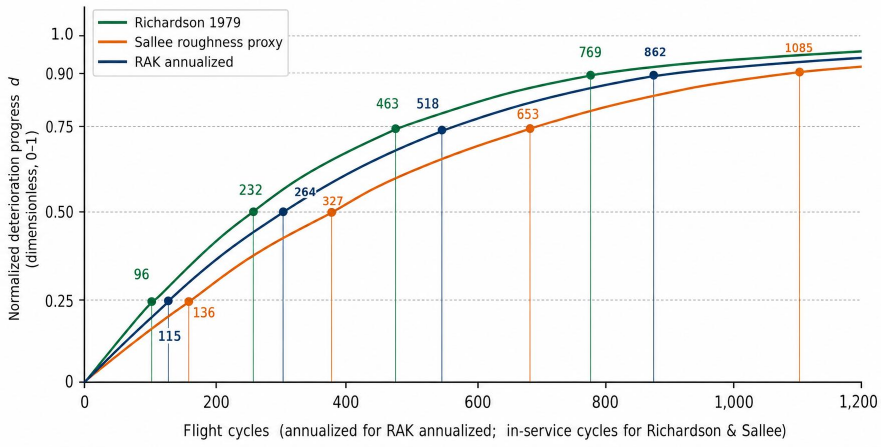}
    \caption{Normalised deterioration $d$ as a function of flight cycles for the Marrakech annualised proxy (RAK airport short code, blue) compared with the two in-service reference datasets. The four milestones $N_{25}$, $N_{50}$, $N_{75}$, and $N_{90}$ are annotated. The horizontal axis represents annualised flight cycles for the model output and in-service cycles for the reference data.}
    \label{fig6}
\end{figure}

In each case, the aircraft operates return flights within the same airport environment, so that both departure and arrival share the same concentration profile. This choice does not imply that real airline operations are restricted to a single environment; it is adopted to isolate the effect of environmental severity on fouling rate while holding other factors constant. Among the three proxies, Marrakech is used for direct comparison with in-service support data, because the two reference datasets \citep{Richardson1979,Sallee1980} reflect long-term fleet averages collected under general operating conditions that are neither extremely clean nor extremely dusty. The first is derived from Richardson et al.~\cite{Richardson1979}, who reported cycle-based HPC deterioration milestones for the JT9D fleet. The second is from Sallee~\cite{Sallee1980}, who documented the JT9D performance deterioration programme with cycle-resolved fouling trends. Since the two in-service reference datasets report long-term deterioration trends without distinguishing seasonal variation, the model output must also be aggregated over seasons to ensure a comparable basis. For each proxy airport, the simulation is run separately for each of the four seasons (DJF, MAM, JJA, and SON), each producing its own set of deterioration milestones. The annualised milestone is then obtained by taking the geometric mean of the four seasonal values for each percentage threshold, which gives equal weight to each season and yields a single representative progression timescale free of seasonal bias. This annualised quantity is what is compared against the in-service reference data throughout this section. Marrakech is selected because it falls between the clean-side proxy (Canary Islands) and the dirty-side proxy (Beijing) within the ordinary-environment family. We define $N_n$ as the number of annualised flight cycles at which the normalised deterioration $d$ reaches $n\,\%$ of its asymptotic value, so that $N_{25}$, $N_{50}$, $N_{75}$, and $N_{90}$ correspond to the 25\,\%, 50\,\%, 75\,\%, and 90\,\% milestones, respectively. At all four deterioration milestones $N_{25}$, $N_{50}$, $N_{75}$, and $N_{90}$, the Marrakech annualised prediction falls between the two in-service reference samples. See \cref{fig6} and \cref{tab:cycle_milestones} for the specific results.

\begin{table}[!htb]
    \centering
    \caption{Comparison of predicted deterioration milestones under the Marrakech annualised proxy with in-service support data.}
    \label{tab:cycle_milestones}
    \setlength{\tabcolsep}{8pt}
    \begin{tabular*}{\linewidth}
    {@{\extracolsep\fill}lllll@{}}
    \toprule
        {\textbf{Sample}} &
        {\textbf{N25}} &
        {\textbf{N50}} &
        {\textbf{N75}} &
        {\textbf{N90}} \\
    \midrule
        Richardson et al.~\cite{Richardson1979} &
        96 & 232 & 463 & 769 \\
        
        Sallee~\cite{Sallee1980} &
        136 & 327 & 653 & 1,085 \\
        
        Marrakech Annualised &
        115 & 264 & 518 & 862 \\
    \bottomrule
    \end{tabular*}
\end{table}

The Marrakech annualised prediction falls between the two reference samples at all four thresholds: closer to the faster-fouling Richardson dataset at early milestones and closer to the slower-fouling Sallee dataset at later milestones. This intermediate positioning is consistent with the moderate environmental severity of the Marrakech proxy. The agreement in both magnitude and timescale provides the basis for using the model as a physics-based explanatory and predictive framework in the following subsections. The comparison is interpreted as the strongest external support rather than engine-specific one-to-one validation, because the engine type, operating history, and environmental exposure records differ between the modelled baseline and the reference fleet data.

\subsection{Environment-to-Performance End-to-End Pipeline Results}\label{sec:closure}

\begin{figure}[!htb]
    \centering
    \includegraphics[width=\columnwidth]{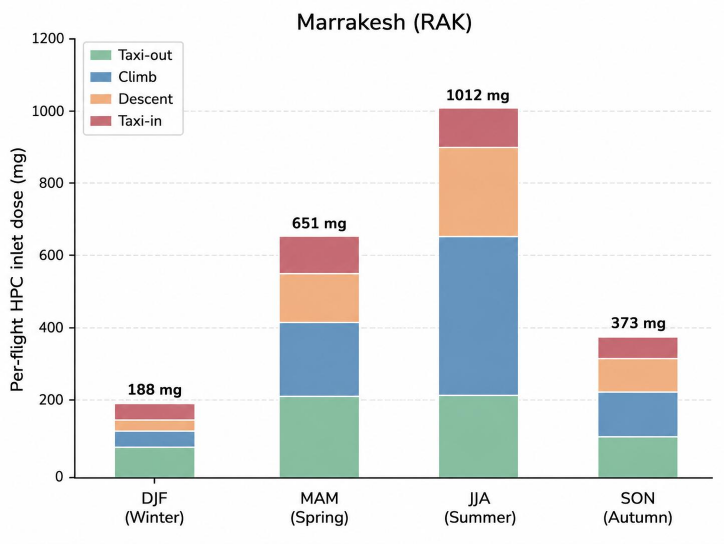}
    \caption{Per-flight HPC inlet particle dose at Marrakech, decomposed by season and flight phase. Each bar represents the total dose from one return flight, with contributions from taxi-out, climb, descent, and taxi-in shown as stacked segments. Summer (JJA) produces the highest dose (1,012\,mg) due to elevated dust activity, while winter (DJF) produces the lowest (188\,mg). These seasonal differences drive the route- and season-dependent fouling behaviour shown in subsequent figures.}
    \label{fig:per_flight_hpc_inlet_dose}
\end{figure}

The proposed framework calculates the overall HPC isentropic efficiency from the environmental conditions and flight profile through a complete forward chain. In this subsection, we present the results of the full pipeline calculation for the Marrakech scenario. The flight profile and altitude-dependent concentration profile described in \cref{sec:scenarios} produce a per-flight HPC inlet dose $C_{\mathrm{HPC,in}}^{(k)}$, calculated using \cref{eq:dose}, that varies with season and flight phase (see \cref{fig:per_flight_hpc_inlet_dose}).

\begin{figure}[!htb]
    \centering
    \includegraphics[width=\columnwidth]{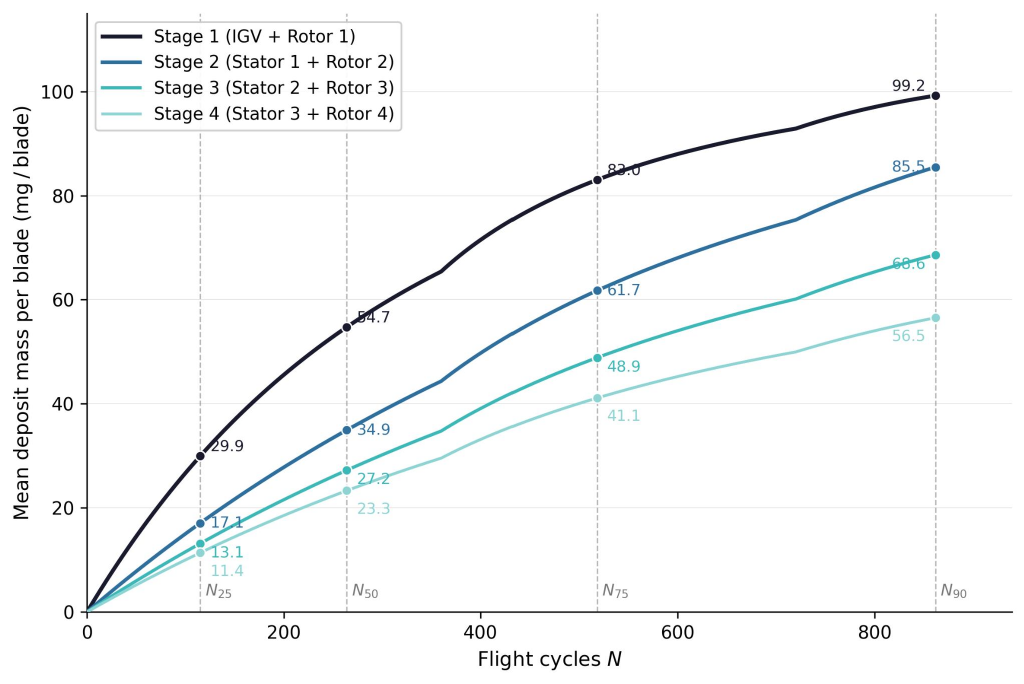}
    \caption{Mean deposit mass per blade $\bar{m}_j(N)$ for the first four HPC stages as a function of flight cycles $N$ under the Marrakech annualised scenario. Values are computed as the geometric mean across the four seasonal windows. Filled circles indicate the four operationally defined milestones $N_{25}$, $N_{50}$, $N_{75}$, and $N_{90}$, with annotated values given in mg per blade.}
    \label{fig:deposition}
\end{figure}

The cumulative dose is distributed across the first four stages through the mass-conserving dose update scheme. Stage~1 captures the largest fraction of the ingested particles, with each downstream stage receiving a progressively reduced residual dose. \cref{fig:deposition} presents the mean deposit mass per blade, $\bar{m}_j(N)$, for Stages~1 through~4 under the Marrakech annualised scenario. The four curves exhibit a strictly monotonic ordering throughout the simulation: $\bar{m}_1(N) > \bar{m}_2(N) > \bar{m}_3(N) > \bar{m}_4(N)$ for all $N$. At $N = 862$ flight cycles, Stage~1 reaches $99.2$\,mg per blade, while Stages~2 through~4 attain $85.5$, $68.6$, and $56.5$\,mg per blade, respectively. This ordering directly reflects the attenuation of particle flux along the axial direction: each successive stage intercepts only the residual dose that has passed through all preceding rows, so that the mean deposit per blade decreases monotonically from inlet to exit.

\begin{figure}[!htb]
    \centering
    \includegraphics[width=\columnwidth]{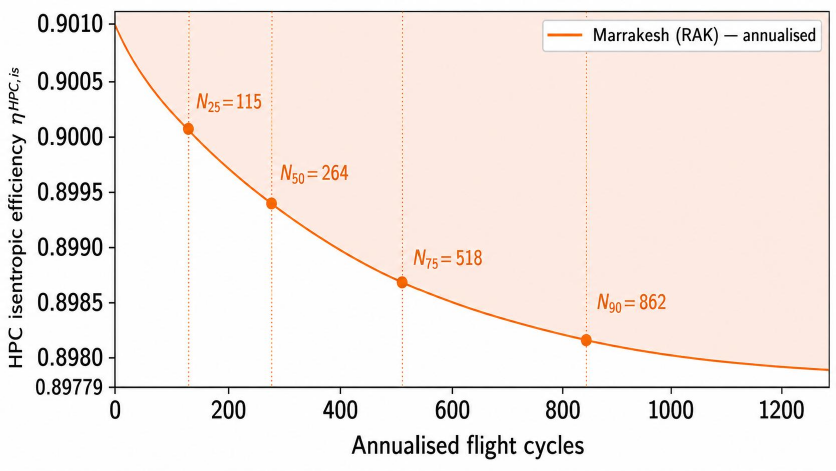}
    \caption{Overall HPC isentropic efficiency deterioration under the Marrakech environment.}
    \label{fig:eta}
\end{figure}

Normalised deposition is assigned to the stage deterioration correction factors $\chi_{\eta,j}$ and $\chi_{\zeta,j}$ through \cref{eq:bridge}, \cref{eq:chi_eta} - \cref{eq:fouled_psi}, which modify the clean-stage coefficients. Stage-by-stage thermodynamic propagation then yields the fouled HPC exit state and the overall isentropic efficiency $\eta'_{\mathrm{HPC,is}}(N)$ (\cref{fig:eta}), calculated using \cref{eq:etaHPC}.

At no point in this chain is the deterioration prescribed in advance. The efficiency loss emerges from the interaction between environmental exposure, deposition capacity, and the stage-stacking thermodynamic propagation. This distinguishes the present approach from models that begin with an assumed level of stage deterioration.

\subsection{Route- and Season-Dependent Fouling Results}\label{sec:routeseason}

Once the model hyperparameters have been optimised (see \cref{tab:params}), the framework is applied to the three proxy environments that span the range of ordinary civil aviation exposure. Each scenario represents a return flight operated entirely within a single airport environment: the aircraft departs and arrives at the same location, so that both the outbound and inbound legs are exposed to the same altitude-dependent concentration profile. The flight profile follows the phase structure described in \cref{sec:scenarios}, comprising taxi-out, climb to 6,000\,m, descent from 6,000\,m, and taxi-in, with airport-specific taxi durations taken from Eurocontrol~\cite{Eurocontrol2024}. The three proxy airports are the Canary Islands (clean-side, ${\approx}9\,\mu\mathrm{g\,m}^{-3}$), Marrakech (central, ${\approx}22\,\mu\mathrm{g\,m}^{-3}$), and Beijing (dirty-side, ${\approx}40\,\mu\mathrm{g\,m}^{-3}$), where the concentrations refer to the altitude-averaged geometric mean over the 0--1,000\,m near-ground layer across all four seasons, derived from the CAMS profiles of Ryder et al.~\cite{Ryder2024}. For each airport, the simulation is run separately under the four seasonal concentration profiles; the resulting deterioration milestones are then combined as a geometric mean across seasons to produce a single annualised trajectory, consistent with the long-term nature of the in-service reference data.

\begin{figure}[!htb]
    \centering
    \includegraphics[width=\columnwidth]{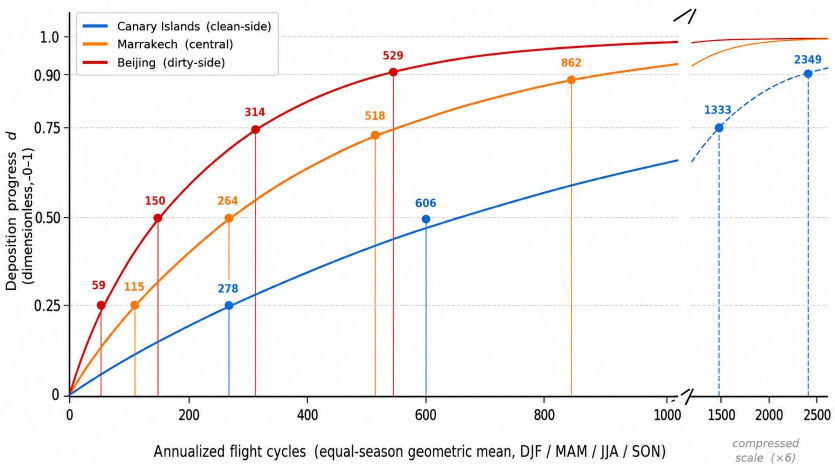}
    \caption{Route-dependent normalised deterioration trajectories under annualised conditions for three proxy environments: Canary Islands (clean-side), Marrakech (central), and Beijing (dirty-side). The three locations are used as environmental proxies to isolate the effect of particle exposure severity on fouling rate, rather than as literal route definitions.}
    \label{fig:proxy_env_deterioration}
\end{figure}

The annualised trajectories in \cref{fig:proxy_env_deterioration} show a clear and consistent ordering across all milestones. Beijing reaches each deterioration milestone first, reflecting its higher ground-level and near-ground particle concentration. Marrakech follows as a representative intermediate environment. The Canary Islands exhibit the slowest deterioration, corresponding to the lowest particle exposure among the three sites. This ordering is preserved at $N_{25}$, $N_{50}$, $N_{75}$, and $N_{90}$, confirming that the route-dependent ranking is robust across the entire deterioration trajectory rather than confined to a single milestone.

Seasonal variability within the Marrakech route is further examined through four full-year simulations, each initiated from a different seasonal starting month. In each simulation, the model advances through all twelve calendar months in sequence: the first three months correspond to the initiating season, after which the inlet particle concentration transitions to that of the subsequent season, thereby altering the per-flight deposition rate. This sequential cycling continues until the full annual flight programme is completed. The four initiating seasons thus produce four distinct deposition trajectories for the same route, reflecting the sensitivity of cumulative fouling to the seasonal concentration profile encountered early in the operating year.

\begin{figure}[!htb]
    \centering
    \includegraphics[width=\columnwidth]{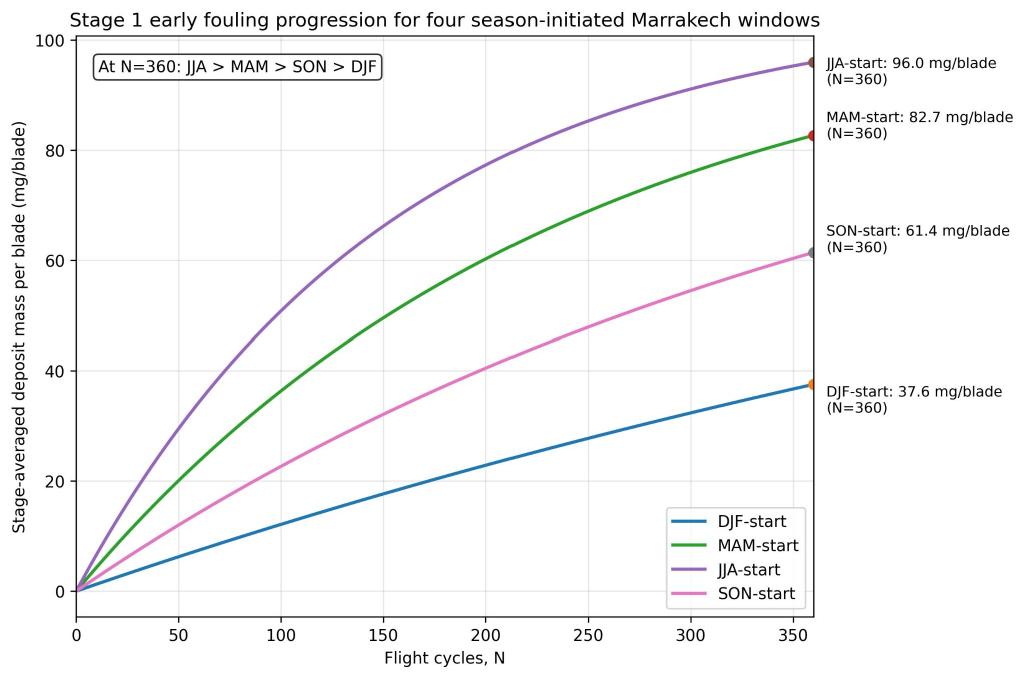}
    \caption{Early Stage~1 deposition progression for the Marrakech case under four season-initiated annual windows. The plotted variable is the stage-averaged deposit mass per blade, computed as the total IGV--Rotor~1 deposited mass divided by the total number of Stage~1 blades. The figure is restricted to the first 360 flight cycles to highlight the effect of the initial seasonal exposure before Stage~1 approaches its common deposition-capacity limit. Endpoint annotations show the corresponding values at $N=360$.}
    \label{fig:seasonal}
\end{figure}

\cref{fig:seasonal} compares the early Stage~1 fouling response for the four Marrakech annual windows initiated in different seasons. The plotted quantity is the stage-averaged deposit mass per blade, calculated from the total deposited mass over the Inlet Guide Vane (IGV)--Rotor~1 pair divided by the total number of Stage~1 blades. The comparison is restricted to the first 360 flight cycles to focus on the initial seasonal forcing before Stage~1 approaches its common deposition-capacity limit. The ordering at the end of this early interval is clear: the JJA-start case reaches the largest deposit mass per blade, followed by the MAM-start, SON-start, and DJF-start cases. This result indicates that the initial seasonal exposure controls the early fouling rate, even though all four annual windows would approach the same Stage~1 capacity over a longer horizon.

\subsection{Physical Interpretation of Deposition and Deterioration Trends}\label{sec:physics}

The model results reproduce several physically expected characteristics of compressor fouling that are well documented in the experimental literature. First, deposition is front-stage dominated: Stage~1 accumulates the largest deposit mass and progresses most rapidly toward saturation, while each subsequent stage captures a smaller share of the residual cumulative dose \citep{Doring2017b,Suman2017review,Zhao2025}. This behaviour arises from the mass-conserving structure of the model, where upstream deposition reduces the particle mass available to downstream stages. Within each stage, stator blades accumulate more deposit mass than rotor blades, consistent with the rotor-to-stator ratio constrained by multistage experimental evidence \citep{Zhao2025,Vulpio2021rotorcraft}.

Second, efficiency deterioration is rapid in the early phase and gradually approaches an asymptotic level. This pattern reflects the exponential saturation of blade deposition capacity: when the surface is clean or lightly fouled, the remaining capacity is large and additional particle mass produces a proportionally larger aerodynamic penalty; as the capacity fills, incremental deposition and the corresponding performance loss diminish. The obtained deterioration trajectory aligns with the exponential saturation form $d \approx 1 - \exp(-N/\tau_{\mathrm{fit}})$ commonly used to describe compressor fouling progression \citep{Doring2017a,Doring2017b,Sallee1980}.

Third, the efficiency loss is driven by the deposition state rather than by the flight-cycle count directly. Two scenarios with the same number of flight cycles but different environmental exposures produce different deposition states and therefore different efficiency losses. This distinction is important because it implies that a simple linear deterioration-per-cycle assumption may underestimate early fouling and overestimate late fouling. The physically traceable intermediate variables at every layer of the model, from dose through deposition progress to correction factors and fouled stage coefficients, make it possible to diagnose where and why different environments produce different deterioration rates. Internal mass-conservation and bounded-state checks are satisfied for all reported simulations.

\section{Conclusion}\label{sec:conclusion}

This work demonstrates that particle-deposition-induced compressor fouling, a problem traditionally studied through isolated blade-row experiments, fixed-condition test rigs, empirical degradation curves, or stand-alone performance correction models, can be reorganised into a systematic forward prediction framework. The proposed framework connects route-, season- and altitude-dependent environmental particle exposure, stage-wise HPC deposition, and overall HPC isentropic efficiency deterioration through a single physically traceable modelling chain. Its contribution lies not in an isolated sub-model, but in the integration of previously fragmented research scales into a continuous information flow: from the operating environment to the particle dose entering the HPC, from the dose to blade-row deposition states, from the deposition states to stage deterioration correction factors, and from the corrected stage coefficients to the overall fouled isentropic efficiency.

The model results confirm four physically expected characteristics of compressor fouling (see \crefrange{fig6}{fig:seasonal}). First, deposition is front-stage dominated, with the first HPC stages accumulating the majority of particle mass while downstream stages receive progressively less cumulative dose. Second, efficiency deterioration is rapid in the early phase of operation and gradually approaches an asymptotic saturation level as the available deposition capacity is consumed. Third, different airports and seasons produce measurably different fouling rates, reflecting the strong dependence of particle exposure on geographic location, seasonal dust activity, and altitude-resolved concentration profiles. Fourth, the predicted deterioration milestones are consistent in magnitude with publicly available in-service support data for HPC deterioration of high-bypass turbofans \citep{Richardson1979,Sallee1980}, providing external evidence that the model operates within an operationally reasonable range. The comparison should be interpreted as support evidence rather than one-to-one engine-specific validation, given the differences in engine type, operating history, and environmental exposure between the modelled baseline and the reference fleet data.

From an academic perspective, the framework bridges local fouling physics and operational deterioration prediction by replacing the conventional approach of prescribing stage deterioration with an environment-driven forward chain that generates deterioration from particle exposure. This shifts compressor fouling analysis from fixed-condition experimental interpretation toward route- and season-resolved operational prediction, and from empirical degradation trends toward a mechanism chain with traceable intermediate variables at every modelling layer. From an industrial perspective, the framework provides a physics-informed basis for engine health monitoring, route- and season-dependent fouling risk assessment, maintenance planning, and compressor washing decisions. Because the modelling architecture depends only on a clean engine baseline, stage geometry, mass-flow profile, and environmental particle data, it is transferable to other civil high-bypass turbofan engines when equivalent input data are available.

Future work will extend the framework to additional engine baselines and incorporate richer airline operational data to further examine the transferability and predictive capability of the model.

\section*{Author Contributions}
Yuyou Zhan: Conceptualisation, data curation, methodology, software implementation, formal analysis, investigation, visualisation, and writing--original draft. Miguel Arana-Catania: Supervision, conceptual and methodological guidance, detailed technical feedback, and writing--review and editing. Neil Dhir: Supervision, methodological advice, technical discussion, and writing--review and editing. Yiguang Li: Supervision, technical advice, research discussion, and writing--review and editing. All authors reviewed and approved the final manuscript.

\section*{Funding Data}
The authors report no funding for this work.

\section*{Conflict of Interest}
There are no conflicts of interest.

\section*{Data Availability Statement}
The data, model code, and other information that support the findings of this study are available from the corresponding author upon reasonable request.


\begin{nomenclature}

\EntryHeading{Roman Symbols}

\entry{$\mathrm{AR}_j$}
{aspect ratio of stage $j$, $h_j/c_j$}

\entry{$a$}
{airport identifier in $C_p(a,s,z)$; also the global
efficiency-prior scaling factor used in Appendix B}

\entry{$b$}
{blade type, stator or rotor; also the global
temperature-coefficient scaling factor used in Appendix B}

\entry{$b_p$}
{pressure-side peak deposit thickness (m)}

\entry{$b_s$}
{suction-side peak deposit thickness (m)}

\entry{$C_{\mathrm{ambient}}$}
{ambient particle mass concentration (kg m$^{-3}$)}

\entry{$\overline{C}_{\mathrm{ann}}$}
{annual geometric-mean particle concentration
($\mu$g m$^{-3}$)}

\entry{$C_p(a,s,z)$}
{particle mass concentration for airport $a$, season $s$, and
altitude $z$ (kg m$^{-3}$)}

\entry{$C_{\mathrm{HPC,in}}^{(k)}$}
{particle mass entering the HPC during flight leg $k$}

\entry{$C_j^{(k)}$}
{particle mass available to stage $j$ during flight leg $k$}

\entry{$C_{\eta,j}$}
{efficiency coefficient used in the fouled-stage mapping,
$1/\eta_{0,j}$}

\entry{$C_{\zeta,j}$}
{temperature-coefficient constant used in the fouled-stage mapping,
$2/\zeta_{0,j}$}

\entry{$c_j$, $c_{\mathrm{geom}}$}
{geometric blade chord length at stage $j$ (m)}

\entry{$c_{\mathrm{eff}}$}
{effective chordwise deposition length, $k_c c_j$ (m)}

\entry{$c_p$}
{specific heat at constant pressure
(J kg$^{-1}$ K$^{-1}$)}

\entry{$c_v$}
{specific heat at constant volume
(J kg$^{-1}$ K$^{-1}$)}

\entry{$d$}
{normalised overall deterioration used to define the cycle
milestones $N_n$}

\entry{$d_j^{(k)}$}
{normalised deterioration driver of stage $j$ after flight leg $k$}

\entry{$f_j^{(k)}$}
{normalised deposition progress of stage $j$ after flight leg $k$}

\entry{$h$}
{specific enthalpy (J kg$^{-1}$)}

\entry{$h_j$, $h_{\mathrm{geom}}$}
{geometric blade span at stage $j$ (m)}

\entry{$h_{\mathrm{eff}}$}
{effective spanwise deposition length, $k_h h_j$ (m)}

\entry{$K$}
{number of HPC stages affected by deposition}

\entry{$k_c$}
{chordwise deposition-coverage factor}

\entry{$k_h$}
{spanwise deposition-coverage factor}

\entry{$m$}
{deposit mass in the Kern--Seaton deposition model}

\entry{$\dot{m}_{\mathrm{bypass}}$}
{bypass-stream mass flow rate (kg s$^{-1}$)}

\entry{$\dot{m}_{\mathrm{core},i}$}
{core mass flow rate in flight-profile segment $i$
(kg s$^{-1}$)}

\entry{$\dot{m}_{\mathrm{total},i}$}
{total engine mass flow rate in flight-profile segment $i$
(kg s$^{-1}$)}

\entry{$m_{s,j}^{(k)}$}
{accepted stator deposit mass per blade at stage $j$ after
flight leg $k$ (g blade$^{-1}$)}

\entry{$m_{r,j}^{(k)}$}
{accepted rotor deposit mass per blade at stage $j$ after
flight leg $k$ (g blade$^{-1}$)}

\entry{$\overline{m}_j(N)$}
{mean deposit mass per blade at stage $j$ after $N$ flight
cycles (mg blade$^{-1}$)}

\entry{$M_{\infty,b,j}$}
{asymptotic single-blade deposit mass for blade type $b$ at
stage $j$ (g blade$^{-1}$)}

\entry{$M_{\infty,j}^{\mathrm{model}}$}
{model-consistent asymptotic total deposit mass of stage $j$ (g)}

\entry{$N$}
{number of flight cycles}

\entry{$N_n$}
{number of flight cycles required to reach $n\%$ of the
asymptotic deterioration}

\entry{$N_1$}
{low-pressure spool speed (\% of reference speed)}

\entry{$n_{s,j}$}
{number of stator blades at stage $j$}

\entry{$n_{r,j}$}
{number of rotor blades at stage $j$}

\entry{$P$}
{thermodynamic pressure (Pa)}

\entry{$p_{t,j}$}
{stagnation pressure at stage station $j$ (Pa)}

\entry{$p_{\mathrm{in},j}$}
{stage inlet pressure (Pa)}

\entry{$p_{\mathrm{out},j}$}
{stage outlet pressure (Pa)}

\entry{$R_0$}
{initial deposition-rate coefficient in the Kern--Seaton model (mass time$^{-1}$)}

\entry{$R_1$}
{deposition-inhibition coefficient in the Kern--Seaton model (time$^{-1}$)}

\entry{$T$}
{thermodynamic temperature (K)}

\entry{$T_{t,j}$}
{stagnation temperature at stage station $j$ (K)}

\entry{$t$}
{time (s)}

\entry{$\Delta t_i$}
{duration of flight-profile segment $i$ (s)}

\entry{$U_j$}
{blade speed at the mean radius of stage $j$ (m s$^{-1}$)}

\entry{$V$}
{thermodynamic volume (m$^3$)}

\entry{$V_{\mathrm{dep},j}$}
{representative deposit volume at stage $j$ (m$^3$)}

\entry{$z$}
{altitude (m)}

\EntryHeading{Greek Symbols}

\entry{$\beta_F$}
{fan particle-transmission factor}

\entry{$\beta_L$}
{LPC-to-HPC particle-transmission factor}

\entry{$\beta^{*}$}
{power-law exponent linking deposition progress to deterioration}

\entry{$\gamma$}
{ratio of specific heats}

\entry{$\gamma_a$}
{effective ratio of specific heats at the mean HPC operating
temperature}

\entry{$\gamma_{rs}$}
{rotor-to-stator deposition ratio}

\entry{$\delta_{\mathrm{layer}}$}
{representative deposit-layer thickness, $b_p+b_s$ (m)}

\entry{$\Delta h_{t,j}$}
{actual clean stage stagnation-enthalpy rise
(J kg$^{-1}$)}

\entry{$\Delta h_{t,s,j}$}
{isentropic clean stage stagnation-enthalpy rise
(J kg$^{-1}$)}

\entry{$\Delta h_{t,j}^{\prime(k)}$}
{actual fouled stagnation-enthalpy rise across stage $j$ after
flight leg $k$ (J kg$^{-1}$)}

\entry{$\Delta M_{s,j}^{(k)}$}
{total stator deposit-mass increment at stage $j$ during
flight leg $k$ (g)}

\entry{$\Delta M_{r,j}^{(k)}$}
{total rotor deposit-mass increment at stage $j$ during
flight leg $k$ (g)}

\entry{$\Delta\eta_{\mathrm{HPC,is}}^{(k)}$}
{change in overall HPC isentropic efficiency after flight leg $k$
relative to the clean baseline}

\entry{$\zeta_{0,j}$}
{clean temperature coefficient of stage $j$}

\entry{$\zeta_j^{\prime *}$}
{normalised fouled temperature coefficient of stage $j$}

\entry{$\zeta_j^{\prime}$}
{actual fouled temperature coefficient of stage $j$ recovered
from the normalised form}

\entry{$\widetilde{\zeta}_j$}
{clean-stage temperature-coefficient shape prior}

\entry{$\eta_{0,j}$}
{clean isentropic efficiency of stage $j$}

\entry{$\eta_j^{\prime *}$}
{normalised fouled isentropic efficiency of stage $j$}

\entry{$\eta_j^{\prime}$}
{actual fouled isentropic efficiency of stage $j$ recovered
from the normalised form}

\entry{$\eta_{\mathrm{HPC,is},0}$}
{clean overall HPC isentropic efficiency}

\entry{$\eta_{\mathrm{HPC,is}}^{\prime(k)}$}
{overall fouled HPC isentropic efficiency after flight leg $k$}

\entry{$\widetilde{\eta}_j$}
{clean-stage efficiency shape prior}

\entry{$\pi_{0,j}$}
{clean pressure ratio of stage $j$}

\entry{$\pi_j^{\prime(k)}$}
{fouled pressure ratio of stage $j$ after flight leg $k$}

\entry{$\rho_{\mathrm{air},i}$}
{air density in flight-profile segment $i$ (kg m$^{-3}$)}

\entry{$\rho_p$}
{deposit material density (kg m$^{-3}$)}

\entry{$\tau$}
{mass-domain deposition time constant (g)}

\entry{$\tau_{\mathrm{fit}}$}
{fitted cycle-domain time constant used to describe asymptotic
deterioration (cycles)}

\entry{$\phi$}
{flow coefficient}

\entry{$\phi^{*}$}
{normalised flow coefficient}

\entry{$\chi_{\eta,j}^{(k)}$}
{efficiency-deterioration correction factor of stage $j$ after
flight leg $k$}

\entry{$\chi_{\eta,\infty,j}$}
{asymptotic efficiency-deterioration correction factor of stage $j$}

\entry{$\chi_{\zeta,j}^{(k)}$}
{loading-deterioration correction factor of stage $j$ after
flight leg $k$}

\entry{$\chi_{\zeta,\infty,j}$}
{asymptotic loading-deterioration correction factor of stage $j$}

\entry{$\psi_{0,j}$}
{clean isentropic pressure-rise coefficient of stage $j$}

\entry{$\psi_j^{\prime *}$}
{normalised fouled isentropic pressure-rise coefficient of stage $j$}

\entry{$\psi_j^{\prime}$}
{actual fouled isentropic pressure-rise coefficient of stage $j$
recovered from the normalised form}

\EntryHeading{Abbreviations}

\entry{BPR}
{bypass ratio}

\entry{CAMS}
{Copernicus Atmosphere Monitoring Service}

\entry{DJF}
{December--February season}

\entry{FPR}
{fan pressure ratio}

\entry{HPC}
{high-pressure compressor}

\entry{IGV}
{inlet guide vane}

\entry{ISA}
{International Standard Atmosphere}

\entry{JJA}
{June--August season}

\entry{LM}
{low-moisture deposition scenario}

\entry{LPC}
{low-pressure compressor}

\entry{MAM}
{March--May season}

\entry{NASA}
{National Aeronautics and Space Administration}

\entry{OPR}
{overall pressure ratio}

\entry{RH}
{relative humidity}

\entry{SON}
{September--November season}

\EntryHeading{Subscripts and Superscripts}

\entry{$0$}
{clean or unfouled condition}

\entry{$3$}
{HPC exit station}

\entry{$25$}
{HPC inlet station}

\entry{$\mathrm{air}$}
{air property}

\entry{$\mathrm{ambient}$}
{ambient condition}

\entry{$\mathrm{ann}$}
{annual or annualised quantity}

\entry{$b$}
{blade type, stator or rotor}

\entry{$\mathrm{bypass}$}
{bypass-stream quantity}

\entry{$\mathrm{capped}$}
{geometrically capped value before application of the
residual-dose constraint}

\entry{$\mathrm{core}$}
{core-flow quantity}

\entry{$\mathrm{dep}$}
{deposition quantity}

\entry{$\mathrm{eff}$}
{effective geometric dimension}

\entry{$F$}
{fan transmission process}

\entry{$\mathrm{fit}$}
{fitted parameter}

\entry{$\mathrm{geom}$}
{geometric dimension}

\entry{$\mathrm{HPC,in}$}
{HPC inlet}

\entry{$i$}
{flight-profile segment index}

\entry{$\mathrm{in}$}
{stage inlet}

\entry{$\mathrm{is}$}
{isentropic quantity}

\entry{$j$}
{HPC stage index}

\entry{$k$}
{flight-leg index}

\entry{$L$}
{LPC-to-HPC transmission process}

\entry{$\mathrm{layer}$}
{deposit-layer quantity}

\entry{$\mathrm{model}$}
{model-consistent quantity}

\entry{$\mathrm{out}$}
{stage outlet}

\entry{$p$}
{particulate or deposit-material quantity; pressure-side quantity
in $b_p$, depending on context}

\entry{$r$}
{rotor}

\entry{$\mathrm{raw}$}
{unconstrained deposition-kernel output}

\entry{$s$}
{stator, isentropic quantity, or season index, depending on context}

\entry{$t$}
{stagnation or total quantity}

\entry{$\mathrm{total}$}
{total engine-flow quantity}

\entry{$\infty$}
{asymptotic or fully saturated condition}

\entry{$'$}
{fouled condition}

\entry{$^{*}$}
{normalised quantity or bridge-exponent designation,
depending on context}

\entry{$\widetilde{\phantom{x}}$}
{shape-prior quantity}

\end{nomenclature}

\appendix

\section{Deposition-Capacity Parameterization}\label{app:depcap}

\cref{tab:depcap} lists the stage-wise blade counts and per-blade deposition capacities used to normalise the accumulated deposited mass into the normalised deposition variable $f_j$. Each stage pairs one stator row with one rotor row; the stage numbering follows the convention Stage~1 = IGV + Rotor~1, Stage~2 = Stator~1 + Rotor~2, and so on. The per-blade saturation mass is computed from the simplified deposition-capacity model described in \cref{sec:depcap}, using the optimal coverage factors $k_c = 0.12$, $k_h = 0.30$, deposit thicknesses $b_p = 0.450$\,mm and $b_s = 0.225$\,mm, and deposit density $\rho_p = 2700$\,kg\,m$^{-3}$.

\begin{table*}[!t]
\caption{Stage-wise blade row geometry and per-blade deposition capacity for the first four HPC stages.}
\label{tab:depcap}
\centering
\small
\setlength{\tabcolsep}{6pt}
\begin{tabular*}{\textwidth}{@{\extracolsep{\fill}}llrrrrrr@{}}
\toprule
Stage & Row & $n$ & $c_{\mathrm{geom}}$ (mm) & $h_{\mathrm{geom}}$ (mm) &
$c_{\mathrm{eff}}$ (mm) & $h_{\mathrm{eff}}$ (mm) & $M_{\infty}$ (g) \\
\midrule
1 & IGV (S)  & 42 & 45.24 & 95.00 & 5.429 & 28.500 & 0.1410 \\
1 & Rotor 1  & 38 & 45.24 & 95.00 & 5.429 & 28.500 & 0.1410 \\
2 & Stator 1 & 82 & 45.24 & 95.00 & 5.429 & 28.500 & 0.1410 \\
2 & Rotor 2  & 53 & 41.57 & 86.25 & 4.988 & 25.875 & 0.1176 \\
3 & Stator 2 & 84 & 41.57 & 86.25 & 4.988 & 25.875 & 0.1176 \\
3 & Rotor 3  & 60 & 37.80 & 77.50 & 4.536 & 23.250 & 0.0961 \\
4 & Stator 3 & 72 & 37.80 & 77.50 & 4.536 & 23.250 & 0.0961 \\
4 & Rotor 4  & 68 & 33.95 & 68.75 & 4.074 & 20.625 & 0.0766 \\
\bottomrule
\end{tabular*}
\end{table*}

\section{Stage-Wise Clean Performance and Deterioration Parameters}\label{app:stageparams}

The clean-stage performance coefficients are constructed by applying dual global scaling factors $a = 0.9055$ and $b = 0.8856$ to the nine-stage shape priors resampled from the ten-stage NASA Energy Efficient Engine dataset \cite{Doring2017b}. The scaling factors enforce simultaneous closure on overall HPC temperature rise and pressure ratio as described in \cref{sec:stacking}. Here, $\gamma_a = 1.38$, corresponding to the effective ratio of specific heats at the mean HPC operating temperature. The resulting nine-stage clean baseline is listed in \cref{tab:cleanbaseline}.

\begin{table*}[!t]
\caption{Nine-stage clean baseline coefficients and stage pressure ratios.}
\label{tab:cleanbaseline}
\centering
\small
\setlength{\tabcolsep}{8pt}
\begin{tabular*}{\textwidth}{@{\extracolsep{\fill}}rrrrrr@{}}
\toprule
Stage & $U_j$ (m s$^{-1}$) & $\zeta_{0,j}$ & $\psi_{0,j}$ & $\eta_{0,j}$ & $\pi_{0,j}$ \\
\midrule
1 & 389.06 & 0.7525 & 0.6875 & 0.9137 & 1.5822 \\
2 & 383.68 & 0.6814 & 0.6402 & 0.9396 & 1.4432 \\
3 & 378.30 & 0.6491 & 0.6100 & 0.9398 & 1.3604 \\
4 & 372.92 & 0.6397 & 0.5983 & 0.9353 & 1.3094 \\
5 & 367.54 & 0.6324 & 0.5868 & 0.9279 & 1.2691 \\
6 & 362.16 & 0.6283 & 0.5797 & 0.9227 & 1.2383 \\
7 & 356.78 & 0.6100 & 0.5603 & 0.9185 & 1.2078 \\
8 & 351.40 & 0.5987 & 0.5450 & 0.9103 & 1.1839 \\
9 & 346.02 & 0.5744 & 0.5091 & 0.8863 & 1.1569 \\
\bottomrule
\end{tabular*}
\end{table*}

For the default scenario reported in the main text (low moisture, $K = 4$), the asymptotic deterioration correction factors are distributed linearly from the first-stage experimental values to unity at stage $K + 1 = 5$, following the spatial model of D\"{o}ring et al. \cite{Doring2017b}. The first-stage values are $\chi_{\eta,\infty,1} = 1.093$ and $\chi_{\zeta,\infty,1} = 1.023$ \cite{Doring2017b}. The resulting stage-wise modifiers are listed in \cref{tab:chi}.

\begin{table*}[!t]
\caption{Asymptotic deterioration correction factors for the default scenario (LM, $K = 4$).}
\label{tab:chi}
\centering
\small
\setlength{\tabcolsep}{10pt}
\begin{tabular*}{0.85\textwidth}{@{\extracolsep{\fill}}rrrl@{}}
\toprule
Stage & $\chi_{\eta,\infty,j}$ & $\chi_{\zeta,\infty,j}$ & Status \\
\midrule
1 & 1.09300 & 1.02300 & Fouled \\
2 & 1.06975 & 1.01725 & Fouled \\
3 & 1.04650 & 1.01150 & Fouled \\
4 & 1.02325 & 1.00575 & Fouled \\
5--9 & 1.00000 & 1.00000 & Clean \\
\bottomrule
\end{tabular*}
\end{table*}

\section{Posterior Candidate Ranges}\label{app:posterior}

\cref{tab:posterior} reports the posterior candidate ranges from which the optimal parameter values used in all main-text simulations were selected. The deposition-layer parameters were calibrated against the cascade data of D\"{o}ring et al.~\cite{Doring2017a} under a fixed ground-level concentration of $48\,\mu\mathrm{g\,m}^{-3}$. The bridge exponent was calibrated against the stage-stacking deterioration results of D\"{o}ring et al.~\cite{Doring2017b}. The complete calibration procedure is described in \cref{sec:calibration_procedure}. Detailed calibration workflows, objective functions, and raw search results are not repeated here.

\begin{table*}[!t]
\caption{Posterior candidate ranges and optimal values for key calibrated parameters.}
\label{tab:posterior}
\centering
\small
\setlength{\tabcolsep}{10pt}
\begin{tabular*}{0.85\textwidth}{@{\extracolsep{\fill}}llr@{}}
\toprule
Parameter & Search range & Optimal value \\
\midrule
$k_c$ & $[0.10,\;0.20]$ & 0.12 \\
$k_h$ & $[0.10,\;1.00]$ & 0.30 \\
$\beta_F$ & $[0.78,\;0.86]$ & 0.78 \\
$\beta_L$ & $[0.77,\;0.984]$ & 0.85 \\
$\gamma_{rs}$ & $[0.44,\;0.52]$ & 0.48 \\
$\beta^{*}$ & $[0.50,\;1.00]$ & 0.94 \\
\bottomrule
\end{tabular*}
\end{table*}

\FloatBarrier

\bibliographystyle{asmejour}
\bibliography{wileyNJD-AMA}

@inproceedings{Richardson1979,
  author = {Richardson, J. H. and Sallee, G. P. and Smakula, F. K.},
  title = {Causes of High Pressure Compressor Deterioration in Service},
  booktitle = {AIAA/SAE/ASME 15th Joint Propulsion Conference},
  venue = {Las Vegas, Nevada, USA},
  eventdate = {June 18--20, 1979},
  number = {AIAA 79-1234},
  year = {1979},
  pages = {1--7},
  doi = {10.2514/6.1979-1234}
}

@techreport{Sallee1980,
  author = "Sallee, G. P.",
  title = "Performance Deterioration Based on Existing (Historical) Data: {JT9D} Jet Engine Diagnostics Program",
  institution = "National Aeronautics and Space Administration",
  number = "NASA CR-135448",
  year = "1978"
}

@article{Doring2017a,
  author = "D{\"o}ring, F. and Staudacher, S. and Koch, C. and Wei{\ss}schuh, M.",
  title = "Modeling Particle Deposition Effects in Aircraft Engine Compressors",
  journal = "ASME Journal of Turbomachinery",
  volume = "139",
  number = "5",
  year = "2017",
  pages = "051003"
}

@inproceedings{Doring2017b,
  author = {D{\"o}ring, F. and Staudacher, S. and Koch, C.},
  title = {Predicting the Temporal Progression of Aircraft Engine Compressor Performance Deterioration due to Particle Deposition},
  booktitle = {ASME Turbo Expo 2017: Turbomachinery Technical Conference and Exposition},
  publisher = {American Society of Mechanical Engineers},
  venue = {Charlotte, North Carolina, USA},
  eventdate = {June 26--30, 2017},
  number = {GT2017-63544},
  year = {2017},
  pages = {V02DT48A007},
  doi = {10.1115/GT2017-63544}
}

@article{Bojdo2020,
  author = "Bojdo, N. and Filippone, A. and Parkes, B. and Clarkson, R.",
  title = "Aircraft Engine Dust Ingestion Following Sand Storms",
  journal = "Aerospace Science and Technology",
  volume = "106",
  year = "2020",
  pages = "106072"
}

@article{Ryder2024,
  author = "Ryder, C. L. and B{\'e}zier, C. and Dacre, H. F. and Clarkson, R. and Amiridis, V. and Marinou, E. and Proestakis, E. and Kipling, Z. and Benedetti, A. and Parrington, M. and R{\'e}my, S. and Vaughan, M.",
  title = "Aircraft Engine Dust Ingestion at Global Airports",
  journal = "Natural Hazards and Earth System Sciences",
  volume = "24",
  year = "2024",
  pages = "2263--2284"
}

@article{Igie2018,
  author = "Igie, U. and Goiricelaya, M. and Nalianda, D. and Minervino, O.",
  title = "Aero Engine Compressor Fouling Effects for Short- and Long-Haul Missions",
  journal = "Proceedings of the Institution of Mechanical Engineers, Part G: Journal of Aerospace Engineering",
  volume = "230",
  number = "7",
  year = "2016",
  pages = "1312--1324"
}

@article{Kurz2012,
  author = "Kurz, R. and Brun, K.",
  title = "Fouling Mechanisms in Axial Compressors",
  journal = "ASME Journal of Engineering for Gas Turbines and Power",
  volume = "134",
  number = "3",
  year = "2012",
  pages = "032401"
}

@article{Bons2010,
  author = "Bons, J. P.",
  title = "A Review of Surface Roughness Effects in Gas Turbines",
  journal = "ASME Journal of Turbomachinery",
  volume = "132",
  number = "2",
  year = "2010",
  pages = "021004"
}

@article{Gbadebo2004,
  author = "Gbadebo, S. A. and Hynes, T. P. and Cumpsty, N. A.",
  title = "Influence of Surface Roughness on Three-Dimensional Separation in Axial Compressors",
  journal = "ASME Journal of Turbomachinery",
  volume = "126",
  number = "4",
  year = "2004",
  pages = "455--463"
}

@article{Suman2017review,
  author = "Suman, A. and Morini, M. and Aldi, N. and Casari, N. and Pinelli, M. and Spina, P. R.",
  title = "A Compressor Fouling Review Based on an Historical Survey of {ASME} Turbo Expo Papers",
  journal = "ASME Journal of Turbomachinery",
  volume = "139",
  number = "4",
  year = "2017",
  pages = "041005"
}

@article{Vulpio2021rotorcraft,
  author = "Vulpio, A. and Suman, A. and Casari, N. and Pinelli, M.",
  title = "Dust Ingestion in a Rotorcraft Engine Compressor: Experimental and Numerical Study of the Fouling Rate",
  journal = "Aerospace",
  volume = "8",
  number = "3",
  year = "2021",
  pages = "81"
}

@article{Bammert1972,
  author = "Bammert, K. and Milsch, R.",
  title = "Das Verhalten der Grenzschichten an rauhen Verdichterschaufeln",
  journal = "Forschung im Ingenieurwesen",
  volume = "38",
  number = "4",
  year = "1972",
  pages = "101--109"
}

@inproceedings{Lakshminarasimha1986,
  author = {Lakshminarasimha, A. N. and Saravanamuttoo, H. I. H.},
  title = {Prediction of Fouled Compressor Performance Using Stage Stacking Techniques},
  booktitle = {Fourth AIAA/ASME Fluid Mechanics, Plasma Dynamics and Lasers Conference},
  venue = {Atlanta, Georgia, USA},
  eventdate = {May 12--14, 1986},
  year = {1986},
  pages = {59--68}
}

@article{Muir1989,
  author = "Muir, D. E. and Saravanamuttoo, H. I. H. and Marshall, D. J.",
  title = "Health Monitoring of Variable Geometry Gas Turbines for the {Canadian} Navy",
  journal = "ASME Journal of Engineering for Gas Turbines and Power",
  volume = "111",
  number = "2",
  year = "1989",
  pages = "244--250"
}

@article{Zaita1998,
  author = {Zaita, A. V. and Buley, G. and Karlsons, G.},
  title = {Performance Deterioration Modeling in Aircraft Gas Turbine Engines},
  journal = {ASME Journal of Engineering for Gas Turbines and Power},
  volume = {120},
  number = {2},
  year = {1998},
  pages = {344--349},
  doi = {10.1115/1.2818128}
}

@article{Yang2014,
  author = "Yang, H. and Xu, H.",
  title = "The New Performance Calculation Method of Fouled Axial Flow Compressor",
  journal = "The Scientific World Journal",
  volume = "2014",
  year = "2014",
  pages = "1--8"
}

@article{Back2012,
  author = "Back, S. C. and Hobson, G. V. and Song, S. J. and Millsaps, K. T.",
  title = "Effects of Reynolds Number and Surface Roughness Magnitude and Location on Compressor Cascade Performance",
  journal = "ASME Journal of Turbomachinery",
  volume = "134",
  number = "5",
  year = "2012",
  pages = "051013"
}

@article{Igie2017,
  author = {Igie, U.},
  title = {Gas Turbine Compressor Fouling and Washing in Power and Aerospace Propulsion},
  journal = {ASME Journal of Engineering for Gas Turbines and Power},
  volume = {139},
  number = {12},
  year = {2017},
  pages = {122602},
  doi = {10.1115/1.4037453}
}

@misc{Halliwell2022,
  author = "Halliwell, I. and Watsek, S. K.",
  title = "A Hybrid-Electric Propulsion System Using Fuselage Boundary Layer Ingestion for a Single-Aisle Commercial Aircraft: Request for Proposal",
  howpublished = "AIAA Foundation Student Design Competition 2022/23",
  year = "2022"
}

@book{Grieb2009,
  author = "Grieb, H.",
  title = "Verdichter f{\"u}r Turbo-Flugtriebwerke",
  publisher = "Springer",
  year = "2009"
}

@manual{Dviation2020,
  author = "{Dviation Training}",
  title = "{CFM56-5B} Engine Borescope Inspection",
  note = "ISS 01, Rev 00, Training Manual",
  year = "2020"
}

@manual{Jiang2004,
  author = "Jiang, W. and Gravelle, G.",
  title = "{CFM56-3} Familiarization Training Manual: Core Major Module",
  organization = "MTU Maintenance Zhuhai Co., Ltd.",
  year = "2004"
}

@article{Turan2022,
  author = "Turan, O.",
  title = "Exergo-Economic Analysis of a {CFM56-7B} Turbofan Engine",
  journal = "Energy",
  volume = "259",
  year = "2022",
  pages = "124936"
}

@book{Farokhi2014,
  author = "Farokhi, S.",
  title = "Aircraft Propulsion",
  publisher = "Wiley",
  edition = "2nd",
  year = "2014"
}

@book{Boyce2011,
  author = "Boyce, M. P.",
  title = "Gas Turbine Engineering Handbook",
  publisher = "Elsevier",
  edition = "4th",
  year = "2011"
}

@techreport{Wu1950,
  author = "Wu, C.-H. and Sinnette, Jr., J. T. and Forrette, R. E.",
  title = "Theoretical Effect of Inlet Hub-Tip-Radius Ratio and Design Specific Mass Flow on Design Performance of Axial-Flow Compressors",
  institution = "National Advisory Committee for Aeronautics",
  number = "NACA TN 2068",
  year = "1950"
}

@article{Mullaney2025,
  author = "Mullaney, D. and others",
  title = "The Effect of Additive Depositional Reprofiling of Compressor Blade Leading Edges on Engine Performance",
  journal = "ASME Journal of Engineering for Gas Turbines and Power",
  volume = "147",
  number = "6",
  year = "2025",
  pages = "061023"
}

@article{Tang2020,
  author = "Tang, J. and others",
  title = "Experimental Study on the Distribution Trends of Fouling on a Compressor Blade",
  journal = "International Journal of Photoenergy",
  volume = "2020",
  year = "2020",
  pages = "8885737"
}

@article{Casari2020,
  author = {Casari, N. and Pinelli, M. and Spina, P. R. and Suman, A. and Vulpio, A.},
  title = {Experimental Assessment of Fouling Effects in a Multistage Axial Compressor},
  journal = {E3S Web of Conferences},
  volume = {197},
  year = {2020},
  pages = {11007},
  doi = {10.1051/e3sconf/202019711007}
}

@article{Vogel2019,
  author = "Vogel, A. and Durant, A. J. and Cassiani, M. and Clarkson, R. J. and Slaby, M. and Diplas, S. and Krüger, K. and Stohl, A.",
  title = "Simulation of Volcanic Ash Ingestion into a Large Aero Engine: Particle--Fan Interactions",
  journal = "ASME Journal of Turbomachinery",
  volume = "141",
  number = "1",
  year = "2019",
  pages = "011010"
}

@article{Vulpio2021timewise,
  author = "Vulpio, A. and Suman, A. and Casari, N. and Pinelli, M. and Kurz, R. and Brun, K.",
  title = "Analysis of Timewise Compressor Fouling Phenomenon on a Multistage Test Compressor: Performance Losses and Particle Adhesion",
  journal = "ASME Journal of Engineering for Gas Turbines and Power",
  volume = "143",
  number = "8",
  year = "2021",
  pages = "081005"
}

@article{Zhao2025,
  author = {Zhao, S. and Yu, X. and Meng, D. and An, G. and Liu, B.},
  title = {Experimental Simulation of Fouling Effects in a Four-Stage Low-Speed Axial Compressor: Fouling Distribution and Performance Changes},
  journal = {Physics of Fluids},
  volume = {37},
  number = {5},
  year = {2025},
  pages = {055143},
  doi = {10.1063/5.0266488}
}

@inproceedings{Tarabrin1998,
  author = {Tarabrin, A. P. and Schurovsky, V. A. and Bodrov, A. I. and Stalder, J.-P.},
  title = {Influence of Axial Compressor Fouling on Gas Turbine Unit Performance Based on Different Schemes and with Different Initial Parameters},
  booktitle = {ASME 1998 International Gas Turbine and Aeroengine Congress and Exhibition},
  publisher = {American Society of Mechanical Engineers},
  venue = {Stockholm, Sweden},
  eventdate = {June 2--5, 1998},
  number = {98-GT-416},
  year = {1998},
  pages = {V004T11A006},
  doi = {10.1115/98-GT-416}
}

@techreport{Eurocontrol2024,
  author = "{Eurocontrol}",
  title = "Taxi Times -- Summer 2024",
  institution = "Eurocontrol",
  year = "2024",
  note = "Available at \url{https://www.eurocontrol.int/publication/taxi-times-summer-2024}"
}

@article{Vulpio2022,
  author  = {Vulpio, Alessandro and others},
  title   = {A Simplified Method for the Deposition Rate Assessment on the Vanes of a Multistage Axial-Flow Compressor},
  journal = {Journal of Turbomachinery},
  volume  = {144},
  number  = {7},
  pages   = {071009},
  year    = {2022}
}

@misc{ISO12103_1_2024,
  author       = {{International Organization for Standardization}},
  title        = {{Road Vehicles---Test Contaminants for Filter Evaluation---Part 1: Arizona Test Dust}},
  howpublished = {ISO 12103-1:2024},
  publisher    = {International Organization for Standardization},
  address      = {Geneva, Switzerland},
  year         = {2024}
}

@article{Syverud2007,
  author = "Syverud, E. and Brekke, O. and Bakken, L. E.",
  title = "Axial Compressor Deterioration Caused by Saltwater Ingestion",
  journal = "ASME Journal of Turbomachinery",
  volume = "129",
  number = "1",
  year = "2007",
  pages = "119--126"
}

@article{AircraftCommerce2008,
  author = {{Aircraft Commerce}},
  title = {{CFM56-7B} Series Specifications},
  journal = {Aircraft Commerce},
  volume = {58},
  year = {2008},
  pages = {10--12}
}

@misc{DeltaTechOps,
  author = "{Delta TechOps}",
  title = "{CFM56-7B} Engine",
  howpublished = "Delta Air Lines Engine Maintenance Services",
  year = "2025",
  note = "Accessed 22 July 2025"
}

@techreport{Ustundag2018,
  author = "{\"U}st{\"u}nda{\u{g}}, V. C. and others",
  title = "Candidate Engines for a Next Generation Supersonic Transport: {ETU--V TULPAR}",
  institution = "TOBB University of Economics and Technology, Dept. of Mechanical Engineering",
  year = "2018"
}

@misc{ScienceDirectJetEngine,
  author       = {{ScienceDirect Topics}},
  title        = {Jet Engine - an Overview},
  year         = {n.d.},
  howpublished = {\url{https://www.sciencedirect.com/topics/earth-and-planetary-sciences/jet-engine}},
  note         = {Accessed: 2026-05-21}
}

@inproceedings{Li2014Pythia,
  author = {Li, Y. G.},
  title = {Training Future Engineers on Gas Turbine Gas Path Diagnostics Using {Pythia}},
  booktitle = {ASME Turbo Expo 2014: Turbine Technical Conference and Exposition},
  publisher = {American Society of Mechanical Engineers},
  venue = {D{\"u}sseldorf, Germany},
  eventdate = {June 16--20, 2014},
  number = {GT2014-25600},
  year = {2014},
  pages = {V006T08A003},
  doi = {10.1115/GT2014-25600}
}

@techreport{ICAO2010,
  author = "{International Civil Aviation Organization}",
  title = "{ICAO} Environmental Report 2010: Aviation and Climate Change",
  institution = "ICAO",
  year = "2010"
}

@article{He2024,
  author = "He, H. and others",
  title = "Numerical Analysis of the Landing and Take-Off Cycle Standard for Supersonic Engines Based on Pollutant Emission Characteristics",
  journal = "Energy",
  volume = "299",
  year = "2024",
  pages = "131424"
}

@inproceedings{Graver2009,
  author = {Graver, B. M. and Frey, H. C.},
  title = {Estimation of Air Carrier Emissions at {Raleigh-Durham} International Airport},
  booktitle = {Proceedings of the 102nd Annual Conference and Exhibition of the Air \& Waste Management Association},
  venue = {Detroit, Michigan, USA},
  eventdate = {June 16--19, 2009},
  number = {2009-A-486-AWMA},
  year = {2009},
  pages = {2502--2516}
}

@inproceedings{Prakash2016,
  author = {Prakash, A.},
  title = {Prediction of {NOx} Emissions Using a Stirred Reactor Modelling Approach for an Aero-Engine with an {RQL} Combustor},
  booktitle = {52nd AIAA/SAE/ASEE Joint Propulsion Conference},
  publisher = {American Institute of Aeronautics and Astronautics},
  venue = {Salt Lake City, Utah, USA},
  eventdate = {July 25--27, 2016},
  number = {AIAA 2016-4501},
  year = {2016},
  pages = {9--18},
  doi = {10.2514/6.2016-4501}
}

\end{document}